\documentclass[pra,aps,superscriptaddress,twocolumn,floatfix]{revtex4-2}

\usepackage{dsfont}

\usepackage{mathtools}
\usepackage{amsmath}
\usepackage{amssymb,mathrsfs}
\usepackage{graphicx}
\usepackage{epstopdf}
\usepackage{mathtext}
\usepackage{yfonts}
\usepackage{bm}

\usepackage{soul}

\usepackage{color}
\usepackage[colorlinks=true,linkcolor=blue]{hyperref}

\newcommand{\br}{{\bm r}}

\begin{document}

\title{Observation of Light Localization at the Edges of Quasicrystal Waveguide Arrays}

\author{S.~K.~Ivanov}
\email[Correspondence email address: ]{sergei.ivanov@uv.es}%
\affiliation{Instituto de Ciencia de los Materiales, Universidad de Valencia, Catedr\'{a}tico J. Beltr\'{a}n, 2, 46980, Paterna, Spain}

\author{S.~A.~Zhuravitskii}
\affiliation{Institute of Spectroscopy, Russian Academy of Sciences, 108840, Troitsk, Moscow, Russia}
\affiliation{Quantum Technology Centre, Faculty of Physics, M. V. Lomonosov Moscow State University, 119991, Moscow, Russia}

\author{N.~S.~Kostyuchenko}
\affiliation{Institute of Spectroscopy, Russian Academy of Sciences, 108840, Troitsk, Moscow, Russia}
\affiliation{Quantum Technology Centre, Faculty of Physics, M. V. Lomonosov Moscow State University, 119991, Moscow, Russia}

\author{N.~N.~Skryabin}
\affiliation{Institute of Spectroscopy, Russian Academy of Sciences, 108840, Troitsk, Moscow, Russia}
\affiliation{Quantum Technology Centre, Faculty of Physics, M. V. Lomonosov Moscow State University, 119991, Moscow, Russia}

\author{I.~V.~Dyakonov}
\affiliation{Quantum Technology Centre, Faculty of Physics, M. V. Lomonosov Moscow State University, 119991, Moscow, Russia}

\author{A.~A.~Kalinkin}
\affiliation{Institute of Spectroscopy, Russian Academy of Sciences, 108840, Troitsk, Moscow, Russia}
\affiliation{Quantum Technology Centre, Faculty of Physics, M. V. Lomonosov Moscow State University, 119991, Moscow, Russia}

\author{S.~P.~Kulik}
\affiliation{Quantum Technology Centre, Faculty of Physics, M. V. Lomonosov Moscow State University, 119991, Moscow, Russia}

\author{V.~O.~Kompanets}
\affiliation{Institute of Spectroscopy, Russian Academy of Sciences, 108840, Troitsk, Moscow, Russia}

\author{S.~V.~Chekalin}
\affiliation{Institute of Spectroscopy, Russian Academy of Sciences, 108840, Troitsk, Moscow, Russia}

\author{S.~Alyatkin}
\affiliation{Hybrid Photonics Laboratory, Skolkovo Institute of Science and Technology, Territory of Innovation Center Skolkovo, Bolshoy Boulevard 30, building 1, 121205, Moscow, Russia.}

\author{A.~Ferrando}
\affiliation{Instituto de Ciencia de los Materiales, Universidad de Valencia, Catedr\'{a}tico J. Beltr\'{a}n, 2, 46980, Paterna, Spain}

\author{Y.~V.~Kartashov}
\affiliation{Institute of Spectroscopy, Russian Academy of Sciences, 108840, Troitsk, Moscow, Russia}

\author{V.~N.~Zadkov}
\affiliation{Institute of Spectroscopy, Russian Academy of Sciences, 108840, Troitsk, Moscow, Russia}
\affiliation{Faculty of Physics, Higher School of Economics, 105066, Moscow, Russia}

\begin{abstract}
Quasicrystals are unique systems that, unlike periodic structures, lack translational symmetry but exhibit long-range order dramatically enriching the system properties. While evolution of light in the bulk of photonic quasicrystals is well studied, experimental evidences of light localization near the edge of truncated photonic quasicrystal structures are practically absent.
{{
In this study, we observe both linear and nonlinear localization of light at the edges of radially cropped quasicrystal waveguide arrays, forming an aperiodic Penrose tiling. Our theoretical analysis reveals that for certain truncation radii, the system exhibits linear eigenstates localized at the edge of the truncated array, whereas for other radii, this localization does not occur, highlighting the significant influence of truncation on edge light localization. Using single-waveguide excitations, we experimentally confirm the presence of localized states in Penrose arrays inscribed by a femtosecond laser and investigate the effects of nonlinearity on these states. Our theoretical findings identify a family of edge solitons, and experimentally, we observe a transition from linear localized states to edge solitons as the power of the input pulse increases.
}}
Our results represent the first experimental demonstration of localization phenomena induced by the selective truncation of quasiperiodic photonic systems. 
\end{abstract}

\maketitle


Quasiperiodic crystals or simply quasicrystals
exhibit long-range order but lack translational symmetry~\cite{Janot-12,Steurer-18}. In contrast to strictly periodic crystals, which can only have certain rotational symmetries,
quasicrystals can have any discrete rotational symmetry, and share the peculiarities of ordered and disordered media, making them complex and appealing for studies. The discovery of quasicrystals in 1982 by Schechtman et al.~\cite{Shechtman-84}
was a milestone showing that aperiodic order can exist in solids. Since then, the study has expanded across various physical systems, enabling controlled investigations of their properties. This research spans from electronic systems~\cite{CollinsWitte-17,KempkesSlot-21} and solid-state physics~\cite{Steurer-04,KamiyaTakeuchi-18}, including
recent applications in twisted bilayer graphene~\cite{AhnMoon18,YaoaWanga-18}, to plasmon polaritons~\cite{VerreAntosiewicz-14}, thin-film ferromagnets~\cite{WatanabeBhat-21}, ultracold atomic systems~\cite{SanchezLewenstein-10,SchreiberHodgman-15,ViebahnSbroscia-19,SbrosciaViebahn-20}, one-dimensional optical semiconductor cavities~\cite{TaneseGurevich-14,BabouxLevy-17,GoblotStrkalj-20}, and photonic systems~\cite{VardenyNahata-13,XuWang-21,WangFuKonotop-24}.

A key aspect of quasicrystalline structures is their influence on wave evolution~\cite{MatsuiAgrawal-07}, localization~\cite{LeviRechtsman-11,Wiersma-13,SegevSilberberg-13} and nontrivial topology~\cite{BandresRechtsman-16,ZhouZhang-19}.
For example, it has been shown that light can be localized in incommensurate (i.e., aperiodic) Moir\'{e} lattices formed by two twisted periodic sublattices~\cite{WangZheng-20,HuangYe-16}.
Recent experiments have shown that linear localization of light can occur
in the bulk of quasicrystals~\cite{WangFuKonotop-24}.
Moreover, the nontrivial geometry of quasicrystals in combination with a nonlinear response of photonic systems leads to many fascinating phenomena in light dynamics~\cite{FreedmanBartal-06,FreedmanLifshitz-07} and soliton formation in the bulk of quasicrystals~\cite{ClausenKivshar-99,AblowitzIlan-06,LawSaxena-10,AblowitzAntar-12,FuWang-20,KartashovYe-21,HuangDong-21}.
Nevertheless, while the linear and nonlinear propagation of light in the bulk of quasicrystals is well studied, it remains largely unexplored at their edges.
Taking into account that the edges of periodic structures repel light, while near-surface localization is possible in disordered lattices, truncated quasicrystals could potentially lead to entirely new scenarios of near-surface localization. {{Beyond photonic systems, these effects have potential implications in
exciton-polaritons in microcavities, aperiodic topological systems, and low-dimensional systems. Furthermore, the current interest in nonlinear atomic systems within aperiodic optical lattices—including Moir\'{e} lattices and quasicrystals—suggests that these findings could be highly relevant for atomic physics as well.}}

In this Letter, we report on unusual linear and nonlinear light localization properties at the edge of a quasiperiodic structure.
We use a laser-fabricated array of waveguides, positioned at the vertices of a Penrose tiling~\cite{Penrose-79}. This array is then truncated using circular masks,
leaving only the waveguides with centers inside the circle of radius $r$. We discover that for certain radii the system exhibits linear eigenmodes localized at the edge of the array, while for other radii this edge localization does not occur. This property is experimentally demonstrated by the direct excitation of edge modes in quasicrystal arrays fabricated in fused silica samples. In contrast, we observe fast diffraction in the bulk of the array.
{{
To interpret this observation, we provide analytical evidence showing that the effect arises from the flatness of angular bands, which may represent a distinct mechanism for edge localization in quasiperiodic systems.
The strong nonlinear response of the implemented optical system allows us to uncover new aspects of this behavior.
}}
Due to the Kerr-type focusing nonlinearity, such edge-localized modes undergo nonlinear bifurcation, giving rise to thresholdless families of edge solitons. They exhibit unique properties stemming from the discreteness of the linear spectrum---a consequence of the array's aperiodicity. This interplay leads to intriguing light behavior in the nonlinear regime.

\begin{figure}[t]
\centering
\includegraphics[width=1.00\linewidth]{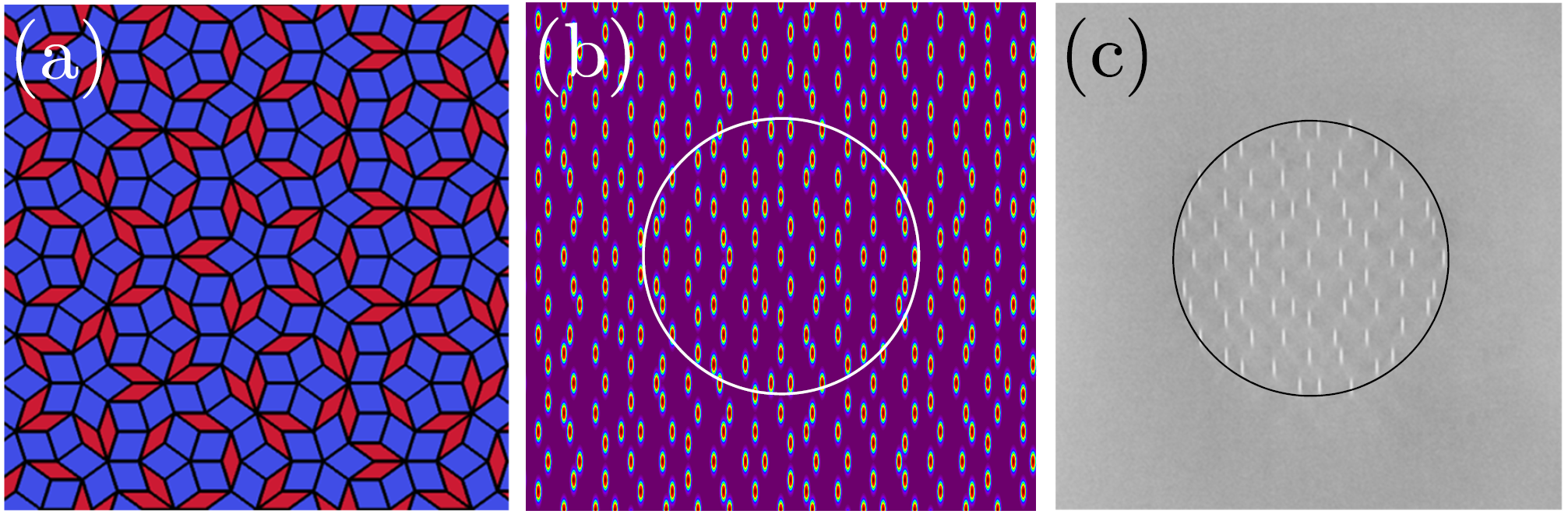}
\caption{
(a) The Penrose P3 tiling 
composed of two types of rhombuses. (b) The quasicrystal waveguide array constructed by placing waveguides at each vertex of the tiling and the truncating circle indicated by the white circumference with radius $r=160$~$\mu\textrm{m}$. (c) Microphotograph of
the waveguide array written with a femtosecond laser.
Here and below arrays are displayed within the window $x,y\in[-300\,\mu \textrm{m},+300\,\mu \textrm{m}]$.}
\label{fig1}
\end{figure}

\begin{figure*}[t]
\centering
\includegraphics[width=1.00\linewidth]{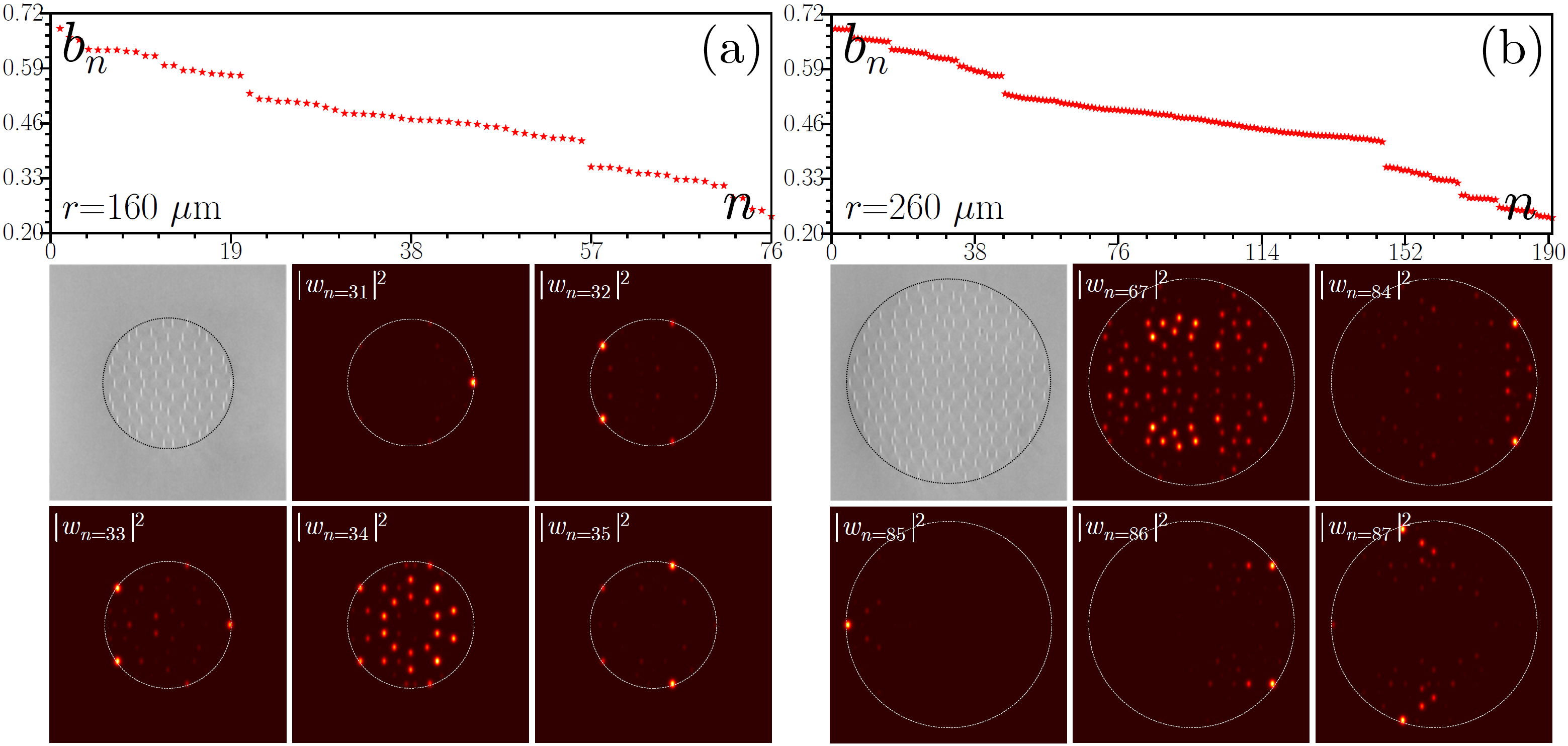}
\caption{
Linear spectra, array profiles, and intensity distributions $|w_n|^2$ in linear modes with different indices $n$ for the truncation radii $r=160$~$\mu\textrm{m}$ (a) and $r=260$~$\mu\textrm{m}$ (b).
}
\label{fig2}
\end{figure*}

\begin{figure}[t]
\centering
\includegraphics[width=1.00\linewidth]{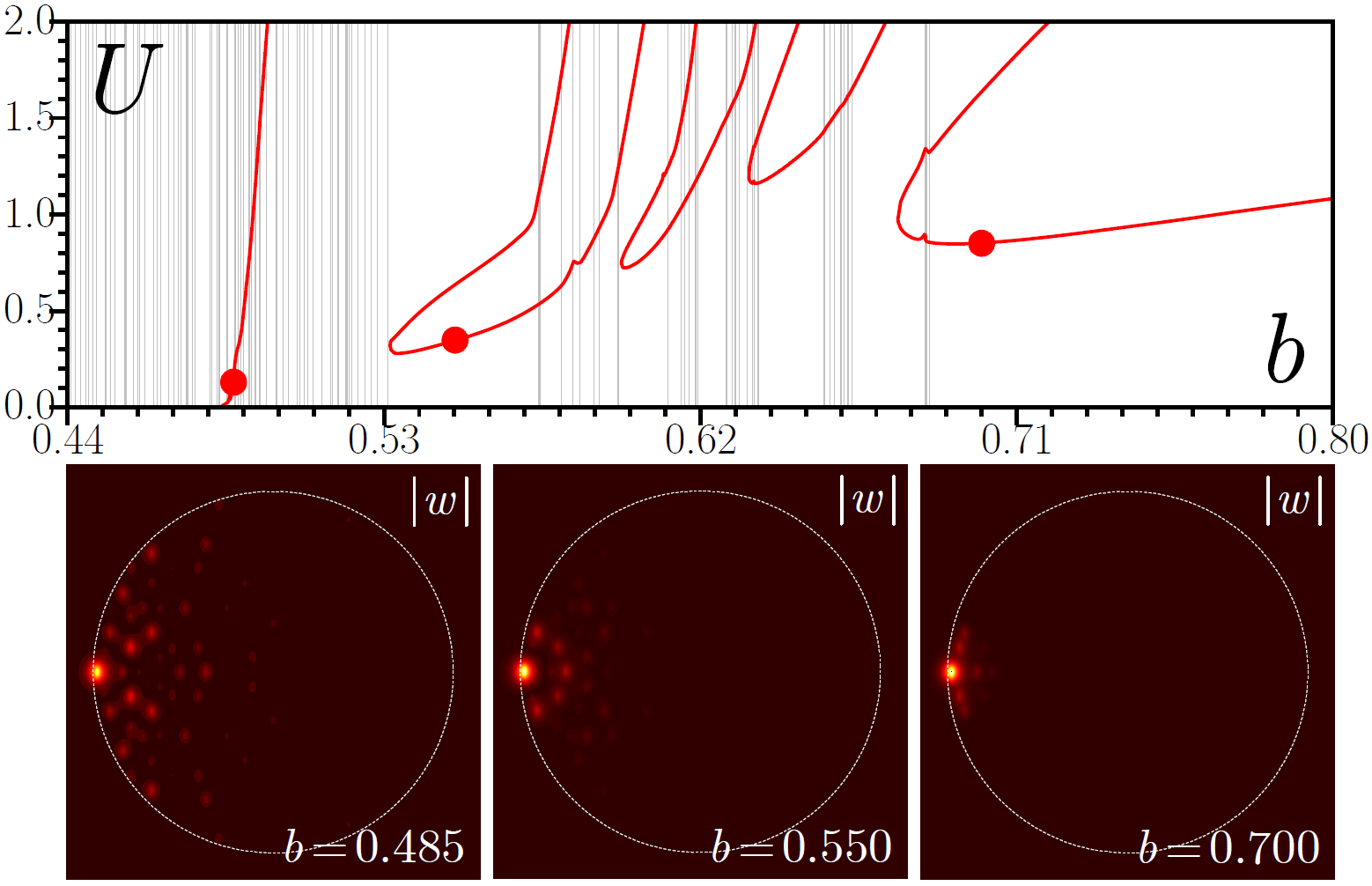}
\caption{
$U(b)$ dependencies illustrating soliton families branching off from the mode localized at the edge of the array with $r=260$~$\mu\textrm{m}$ and its ``continuations'' in other minigaps. Vertical gray lines indicate the eigenvalues of the linear modes of the quasiperiodic array. Field modulus distributions $|w|$ in solitons corresponding to the points in the $U(b)$ curves.
}
\label{fig3}
\end{figure}


\begin{figure}[t]
\centering
\includegraphics[width=1.00\linewidth]{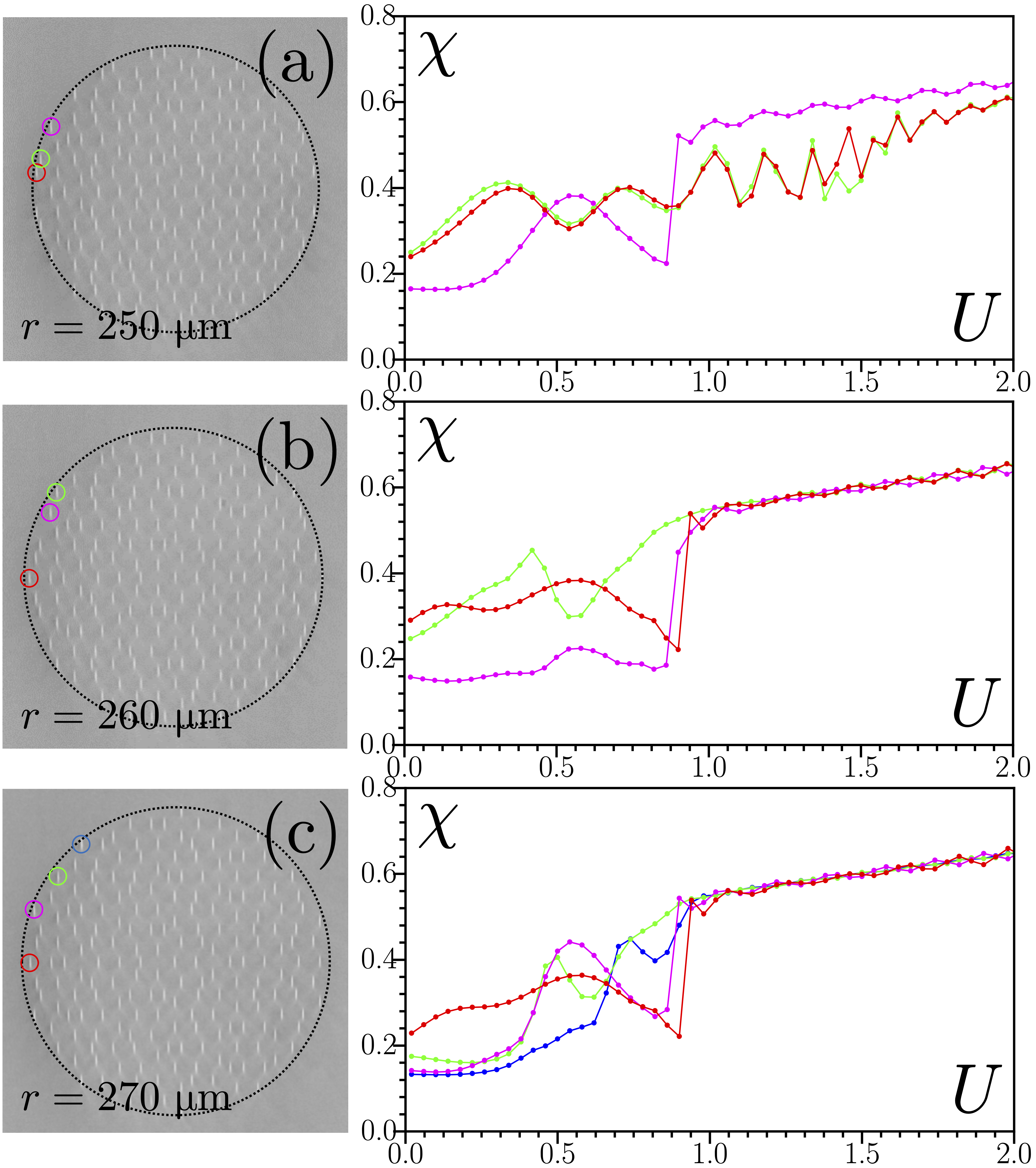}
\caption{
Photographs of the arrays written with the femtosecond laser (left column) and the theoretically calculated output form factor $\chi$ against the input power $U$ (right column) for three truncation radii: $r=250$~$\mu\textrm{m}$~(a), $r=260$~$\mu\textrm{m}$~(b), and $r=270$~$\mu\textrm{m}$~(c). The color coding of the lines corresponds to the circles that indicate the excited guides.
}
\label{fig4}
\end{figure}

\begin{figure*}[t]
\centering
\includegraphics[width=1.00\linewidth]{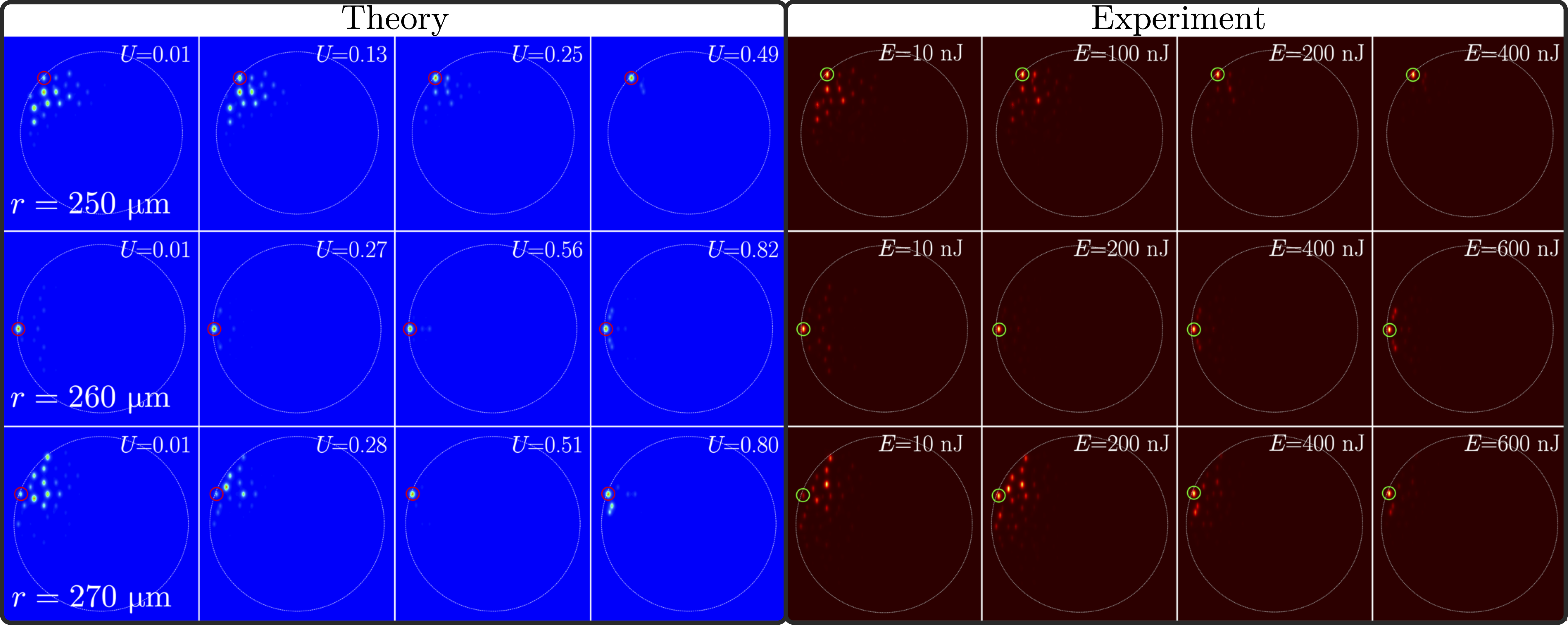}
\caption{
Theoretical (blue background) and experimental (maroon background) intensity distributions for different input pulse energies for $r=250$~$\mu\textrm{m}$ (top row), $r=260$~$\mu\textrm{m}$ (middle row), and $r=270$~$\mu\textrm{m}$ (bottom row). Photomicrographs of the waveguide arrangements for these radii are shown in Fig.~\ref{fig4}. The red and green circles indicate the excited waveguide.
{{
The dimensionless power of $U=0.1$ approximately corresponds to the pulse energy $E=75$~nJ.
}}
}
\label{fig5}
\end{figure*}

The propagation of light beams with dimensionless amplitude $\psi(\br,z)$ in waveguide arrays inscribed in a focusing cubic medium is described by the nonlinear Schr\"{o}dinger equation:
\begin{align}  
\label{eqNLS}
    i\frac{\partial\psi}{\partial z} = -\frac{1}{2}\nabla^2\psi -\mathcal{R}(\br)\psi - |\psi|^2\psi.
\end{align}
Here $\nabla=(\partial/\partial x,\partial/\partial y)$, the coordinates $\br=(x,y)$ are normalized to the characteristic transversal scale $r_0=10$~$\mu \textrm{m}$, the propagation distance $z$ is normalized to the diffraction length $kr_0^2$, $k = 2\pi n_0/\lambda$, $\lambda=800$~$\textrm{nm}$, and $n_0\approx 1.45$ is the unperturbed refractive index.

Here we consider quasicrystal arrays based on the Penrose P3 tiling, which consists of two types of rhombuses with equal side lengths $d$ but different angles [see Fig.~\ref{fig1}(a)]. This structure has a fivefold rotational symmetry.
In this study, we place the centers of the waveguides at the vertices of the tiling. The function $\mathcal{R}(\br) = p\sum_{n,m} e^{-\left[(x-x_n)^2/w_x^2+(y-y_m)^2/w_y^2\right]}$ describes single-mode Gaussian waveguides, where $x_n$ and $y_m$ are the coordinates of the vertices of the Penrose tiling. The exemplary profile of the waveguide array
is shown in Fig.~\ref{fig1}(b). Since we are interested in the effects of truncation on the modes of an infinite array, we
spatially crop the Penrose array, using a circular mask of radius $r$ centered at the origin, as schematically shown with white circumference in Fig.~\ref{fig1}(b), corresponding to a radius of $r=160$~$\mu\textrm{m}$. Waveguides with centers outside the mask are removed, leaving a finite array with $N$ waveguides inside. Figure~\ref{fig1}(c) shows a microscopic image of one of the structures (with $N=76$ waveguides) fabricated for the experiment. As we conclusively demonstrate below, aperiodic structures of different sizes ($r$) reveal different localization properties. The minimum distance between the waveguides is $d=23$~$\mu\textrm{m}$ and $w_x=2.4$~$\mu\textrm{m}$, $w_y=9.6$~$\mu\textrm{m}$ are the widths of each waveguide (they are elliptical due to the fs-writing method) along the $x$ and $y$ axes. The depth of the waveguides is given by $p = k^2r_0^2\delta n/n_0$, where $\delta n$ is the refractive index contrast. Here we set $p = 4.85$, which corresponds to $\delta n\approx5.4\cdot10^{-4}$.

We first {{omit the nonlinear term in (\ref{eqNLS}) and}} focus on the analysis of the \emph{linear} eigenmodes $\psi(\br,z)=w_n(\br) e^{ib_nz}$ of a truncated quasiperiodic array, where $b_n$ is the propagation constant of mode with index $n$ (only the discrete spectrum at $b_n>0$ is considered). We have found that localized edge modes only occur for certain truncation radii $r$. Two examples of arrays with localized edge modes are shown in Fig.~\ref{fig2} for radii of $160$~$\mu\textrm{m}$ [Fig.~\ref{fig2}(a)] and $260$~$\mu\textrm{m}$ [Fig.~\ref{fig2}(b)] together with spectra and modal profiles. The dashed black lines indicate the radii of truncation. The propagation constants are sorted in descending order. Although the array is aperiodic, the formation of several groups of eigenvalues (i.e., quasi-bands) is noticeable. One of these quasi-bands contains modes that are localized at the edges of the array. For example, for a radius of $r=160$~$\mu\textrm{m}$ (producing an array with $N=76$ waveguides) there are five edge modes with the indices $n=31$, $32$, $33$, $35$, $36$. The number and properties of these edge modes are strongly linked to the discrete rotational symmetry of the system. A hallmark of strong localization at the edge is the appearance of angular flat bands with
$b$ approximately equal to that of the fundamental mode of an isolated waveguide
(see Appendix),
although in our system, waveguides, where the edge modes are predominantly localized, are not isolated.
Four examples of the intensity profiles of such edge modes are shown in Fig.~\ref{fig2}(a). For contrast, a delocalized bulk mode with index $n=34$ is also shown. It is noteworthy that a slight change in the truncation radius, adding or removing a layer of waveguides, causes these edge modes to disappear. However, for a larger radius of $260$~$\mu\textrm{m}$ with $N=191$ waveguides [Fig.~\ref{fig2}(b)], we also found modes localized at the edge. These modes have the indices $n=84$, $85$, $86$, $87$, $88$. Examples of the intensity profiles of four of these modes are shown in Fig.~\ref{fig2} (b). The profile of a delocalized mode with index $n=67$ is also included. {{To confirm that edge-localized modes are not artifacts of specific parameter choices, we verified their persistence for other values of $d$, while preserving the array’s outer boundary structure (see Appendix).}}

To study the bifurcation of the family of edge \emph{solitons} from linear edge states, we now consider Eq.~(\ref{eqNLS}) with cubic nonlinearity included. In the presence of nonlinearity, localized edge modes in our arrays give rise to the families of solitons. Such solitons have the form $\psi = w(\br)e^{ibz}$,
where now the propagation constant $b$ parameterizes the family of solitons and determines their power $U = \int |\psi|^2 d^2\br$. Figure~\ref{fig3} shows the nonlinear family for $r=260$~$\mu\textrm{m}$. The vertical gray lines
correspond to propagation constants of the linear modes. As the figure shows, the family is thresholdless: When the propagation constant tends toward that of the linear edge mode ($b_{85}=0.483$), the power $U$ vanishes and the shape of the soliton approaches the shape of this linear edge mode. {{It can be shown~\cite{Konotop2024} that near the linear limit $(b\to b_n)$ the following formula can be derived $U(b)= (b-b_n)/\chi^2$, where $\chi^2=\int|w_n|^4d^2\br$, which accurately describes the actual $U(b)$ dependence near the bifurcation point from the linear state; see the comparison in Appendix.}}
Since the linear edge mode is in the quasi-band, increasing $b$ leads to coupling of the soliton with bulk modes, resulting in the soliton acquiring a long tail in the array, as seen in the soliton profile for $b=0.485$. Given the complexity and discreteness of the linear spectrum, coupling with different modes leads to multiple branches of the nonlinear family,
which results from the aperiodicity of the structure. In Fig.~\ref{fig3} we have omitted some hybrid families combining edge and bulk states and instead focused on the simplest solitons localized at the edge of the array. Despite the complex $U(b)$ dependence, these solitons remain predominantly localized near the edges for almost the entire family (see, e.g., soliton profile for $b=0.55$). For large nonlinear propagation constants $b$ we obtain solitons belonging to a semi-infinite gap of the linear spectrum (see $b=0.7$). {{Additionally, we employed a variational approach to analytically describe the $U(b)$ dependence for sufficiently high powers. Further details regarding this approximation and a discussion of stability can be found in Appendix. Solitons bifurcating from the linear modes of the quasicrystal structure represent simple (non-excited) nonlinear states whose stability properties are adequately characterized by the Vakhitov-Kolokolov stability criterion \cite{VakhitovKolokolov-73}. This suggests that collapse, which is ubiquitous in uniform cubic media \cite{Berge1998}, cannot occur for these stable nonlinear states. This is in agreement with previous findings that periodic and aperiodic potentials do suppress instabilities for certain types of solitons \cite{AcevesTuritsyn1994, AcevesTuritsyn1995, KivsharPelinovsky2000, Kartashov-19, Malomed2024, ZhangKartashov2024}.}}


In the experiment, we inscribed Penrose arrays into $10$~cm long fused silica glass using the femtosecond laser writing technique (see~Appendix).
Microscopic images of the fabricated quasiperiodic arrays are shown in Fig.~\ref{fig4}
for three truncation radii of $250$, $260$, and $270$~$\mu\textrm{m}$. The length of our sample corresponds to a dimensionless length of $z \approx 88$. We performed a theoretical analysis of the output distributions for the single-site excitation,
which is shown in Fig.~\ref{fig4}. The colored circles indicate the positions for single-site excitation in different waveguides at the edge of the quasicrystal. This figure shows the dependence of the output form factors $\chi=U^{-1}\left[\int|\psi|^4d^2\br\right]^{1/2}$ on the power of the initial pulse for the excitation of different waveguides at the edge. Higher $\chi$ values indicate better localization. In the linear regime, the highest form factor $\chi=0.29$ occurs when the waveguide marked with a red circle is excited in an array with a radius of $260$~$\mu\textrm{m}$. This corresponds to the excitation of the edge mode shown in Fig.~\ref{fig2}(b). In other cases, the form factor is lower. For example, for a truncation radius of $250$~$\mu\textrm{m}$, $\chi$ ranges from $0.16$ for the magenta circle to $0.23$ for the green circle, indicating the simultaneous excitation of several bulk modes.
This is supported by calculated weights of the different linear modes involved in the process for different types of excitation (see Appendix).
Strong light localization corresponds to the excitation of the localized edge mode with the highest weight, while rapid diffraction results from the simultaneous excitation of multiple modes. It is worth noting that as the propagation length $z$ increases, the differences between $\chi$ for different excitations become more pronounced. As the power of the initial state increases, the curves exhibit non-monotonic behavior, indicating coupling with different system modes. At high power $U\geq1$, the light is strongly localized in the excited waveguide.
{{To reliably highlight the effects of exciting different waveguides, we also performed numerical simulations for structures with a longitudinal length $z$ significantly larger than that of the experimental sample. These simulations illustrate how the coordinates of the integral beam center evolve during propagation along the sample (see~Appendix). When the waveguide marked in red in Fig.~\ref{fig4}(b) is excited, light remains predominantly localized within the excited waveguide, with the beam center trajectory located in the close vicinity of this channel.
For other excited waveguides, both in the bulk and at the edge of array, the trajectories of beam center are much longer, indicating on strong delocalization of the initial single-channel excitations (see~Appendix for details).
}}

To probe the localization properties, we have performed an experiment with $280$~fs pulses of variable energy $E$ using a $1$~kHz femtosecond Ti:sapphire laser {at 800 nm central wavelength (normal dispersion regime, see~Appendix for details).} Some illustrative examples of output distributions for the excitation of the outermost waveguides are shown in Fig.~\ref{fig5} for three truncation radii: $250$, $260$ and $270$~$\mu\textrm{m}$. The excitation of the edge waveguides in the structures with $r=250$ and $270$~$\mu\textrm{m}$ leads to strong diffraction in the linear regime. This is because single-site excitation of the selected waveguide simultaneously populates many linear modes of the system with comparable weights, most of which are delocalized (see~Appendix). A similar broadening can also be observed for the excitation of central waveguides. As the energy of the input pulse increases, there is a gradual contraction of the output pattern and eventually the formation of a well-localized
soliton. In contrast, excitation of the waveguide at the edge of the quasicrystal for a radius of $260$~$\mu\textrm{m}$ leads to virtually no broadening. At low energies, the output pattern is very similar to the calculated linear edge modes from Fig.~\ref{fig2}(b). Simulations over much larger distances, which drastically exceed our sample length, also show no broadening of the pattern.
Increasing the energy leads to coupling with other modes, resulting in increased tails in neighboring waveguides. However, a further increase in energy leads to the appearance of a strongly localized edge soliton. We have also experimentally investigated the excitation of waveguides located at the edge of the arrays for other radii and found a strong localization for a truncation radius of $160$~$\mu\textrm{m}$, which is in complete agreement with the localization properties shown in Fig.~\ref{fig2}, while no localization is observed for radii of $155$ and $170$~$\mu\textrm{m}$ (see~Appendix) for the experimental output distributions).


In summary, for the first time to our knowledge, we have observed a significant dependence of light localization in both linear and nonlinear regimes on the system spatial size and the excitation position within a quasiperiodic array based on radially cropped Penrose tiling.
{{Our results on experimental platform allowing to create aperiodic arrays with clearly defined edges open the avenue for exploration of unconventional localization phenomena near edges of aperiodic structures with long-range order with potential implications in solid-state physics, low-dimensional systems, physics of matter waves, and polariton systems. Our advanced platform, fabricated using femtosecond laser inscription, enables experiments spanning linear to strongly nonlinear regimes.}}

~

\begin{acknowledgments}
We are grateful to Vladimir~V.~Konotop for useful discussions.
\end{acknowledgments}

\textbf{Funding.}
This project has received funding from the Ministerio de Ciencia e Innovaci\'{o}n (PID2020-120484RBI00 and COMCUANTICA/009 (with funding from European Union NextGenerationEUPRTR-C17.I1)), the Generalitat Valenciana PROMETEO/2021/082,
{{ and
the European Union through the  Program Fondo Social Europeo Plus 2021-2027(FSE+) of the Valencian Community (Generalitat Valenciana CIAPOS/2023/329)}}. It was also supported in part by research project FFUU-2024-0003 of the Institute of Spectroscopy of the RAS and by the Russian Science Foundation (Grant 24-12-00167). S. A. Zhuravitskii acknowledges support by the Foundation for the Advancement of Theoretical Physics and Mathematics ``BASIS'' (22-2-2-26-1).


\appendix

\renewcommand{\thefigure}{A\arabic{figure}}
\setcounter{figure}{0}

\section{Femtosecond laser written structures and Experimental excitation of waveguides}\label{sec1}

For the experimental observation of localization in quasiperiodic structures, we use waveguide arrays based on Penrose tiling inscribed inside $10$~$\rm cm$ long fused silica glass samples by focused (using an aspherical lens with $\rm NA = 0.3$) femtosecond laser pulses (wavelength $515$~$\rm nm$, pulse duration $230$~$\rm fs$, pulse energy $320$~$\rm nJ$, repetition rate $1$~$\rm MHz$). During the inscription process, the sample was translated relative to the focus at a constant speed of $1$~$\rm mm/s$ using a high-precision air-bearing positioner (Aerotech), resulting in the inscription of sets of parallel waveguides with controllable spacing between them. The contrast of the refractive index change in such waveguides is about $\delta n\sim 5.4\cdot10^{-4}$, i.e., they are single-mode with the mode field diameters $\sim 14.3$~$\mu\rm m$ by $21.3$~$\mu\rm m$. The waveguides show propagation losses of less than $0.3$~dB/cm at $\lambda = 800$~nm.

In experiments, we have used single-waveguide excitations with fs pulses of variable energy $E$ from a $1$~kHz Ti:sapphire laser with a center wavelength of $800$~nm. Short pulses with a duration of $40$~fs and a broad spectrum from a Spitfire HP regenerative amplifier system (Spectra Physics) first pass through an active beam position stabilization system (Avesta) and an attenuator and then fed into a $4$f single grating variable-slit stretcher-compressor. The spectra of such pulses are narrowed by a slit to $5$~nm, which corresponds to a pulse duration of $280$~fs. This increase in pulse duration makes it possible to prevent optical collapse and strong spectral broadening during pulse propagation in the waveguides, i.e., it allows the temporal effects to be neglected. The pulses after the stretcher-compressor were focused to excite single waveguides and the output intensity distributions
were recorded with a Kiralux CMOS camera (Thorlabs). The input peak power in the waveguide (for each pulse in the $1$~kHz sequence) was defined as the ratio of the input pulse energy $E$ to the pulse duration $\tau=280$~fs. Taking into account the losses for matching with the focusing lens, the input power can be evaluated as $2.5$~kW for each $1$~nJ. The maximum excitation energy used in experiments was $E = 600$~nJ corresponds to peak power of $1.5$~MW.


\begin{figure*}[t]
\centering
\includegraphics[width=1.00\linewidth]{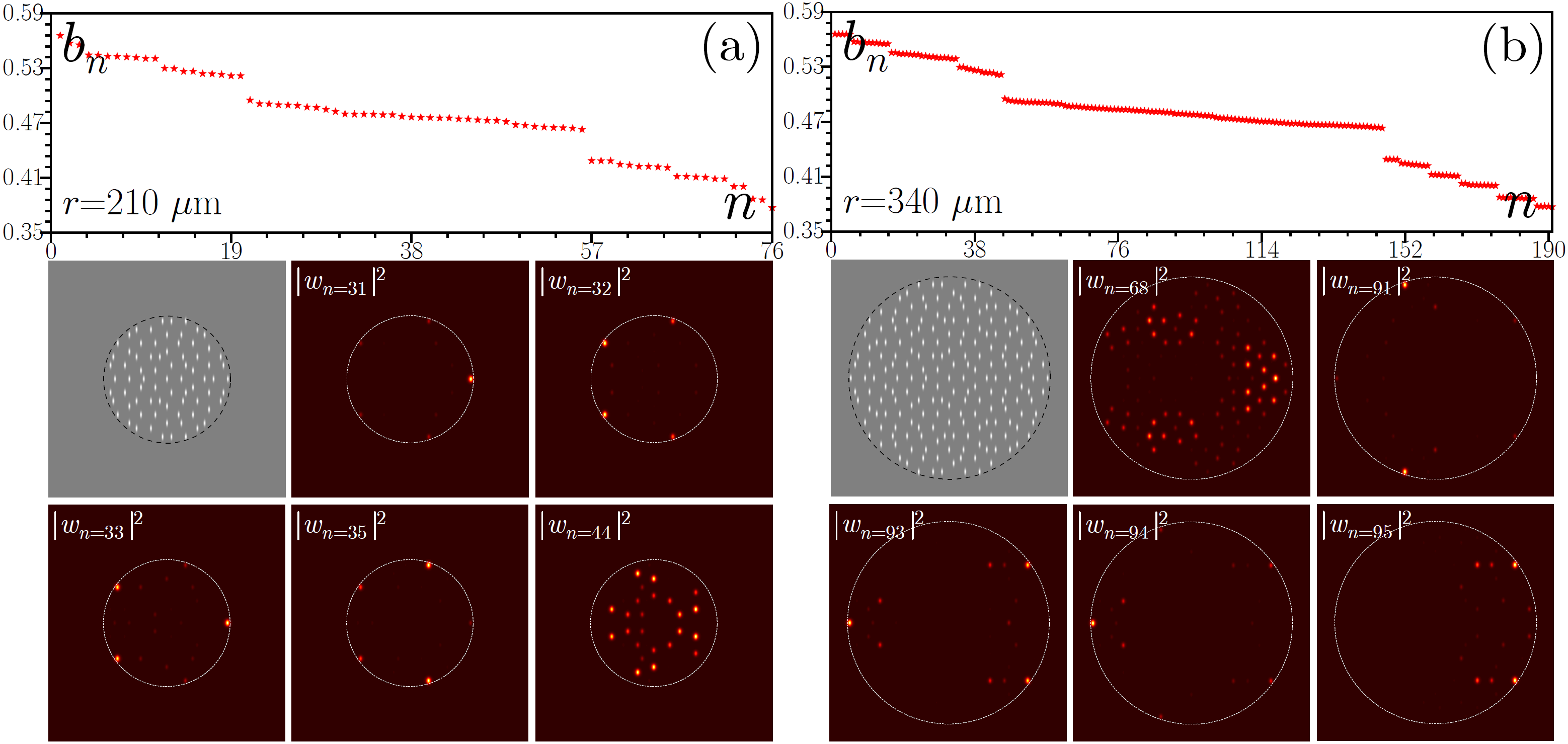}
\caption{
{{
Linear spectra, array profiles and intensity distributions $|w_n|^2$ in linear modes with different indices $n$ in quasicrystal structure with enlarged minimal distance between the waveguides $d=30$~$\mu\textrm{m}$, for the truncation radii $r=210$~$\mu\textrm{m}$ (a) and $r=340$~$\mu\textrm{m}$ (b). The arrays and modes are shown within the window $x,y\in[-400\,\mu \textrm{m},+400\,\mu \textrm{m}]$.
}}
}
\label{figS1}
\end{figure*}

\begin{figure}[t]
\centering
\includegraphics[width=1\linewidth]{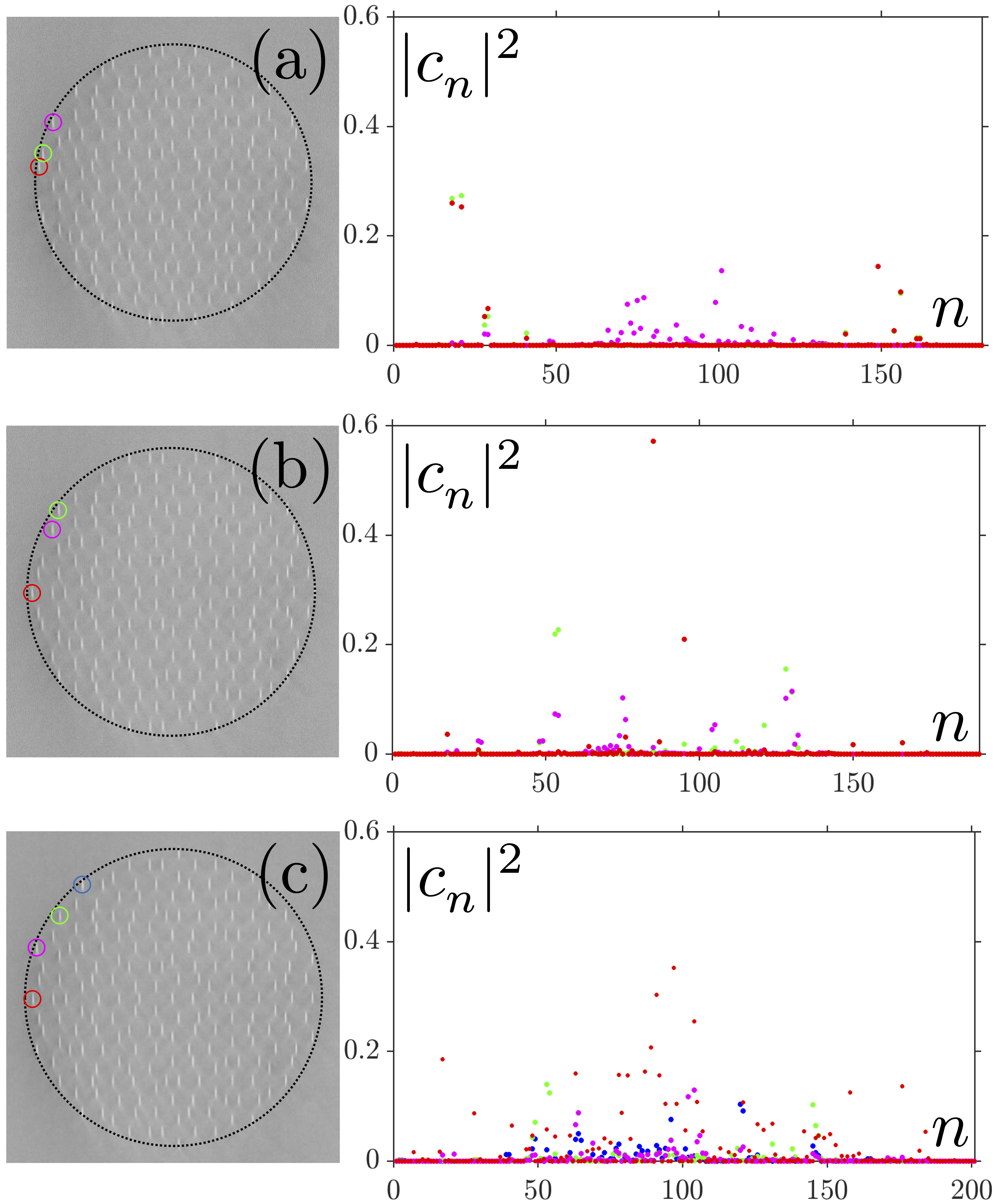}
\caption{
Photographs of the arrays written with the femtosecond laser (left column) and theoretically calculated weights $|c_n |^2$ of all linearly excited modes (right column) for a single-site excitation for three radii of truncation: $r=250$~$\mu\textrm{m}$~(a), $r=260$~$\mu\textrm{m}$~(b) and $r=270$~$\mu\textrm{m}$~(c). The color coding of the lines corresponds to the circles that indicate the excited guides.}
\label{figS2}
\end{figure}

{{

\section{Linear edge modes for larger array scale}\label{sec2}

To highlight the generality of our results, we demonstrate the existence of linear edge modes in truncated quasicrystal waveguide array with similar internal structure as arrays described in the main text, but with increased spacing between waveguides. Specifically, we increased the minimum spacing between waveguides to $d=3$. The linear spectra, array profiles, and examples of modes for this case are presented in Fig.~\ref{figS1}. As in the smaller-scale structures described in the main text, localized edge modes occur only for certain truncation radii  $r$. In structure with increased scale, localized edge modes appear when the radius of truncating ring is increased accordingly, leading to qualitatively similar structure of the boundaries as in arrays shown in Fig.~2 of the main text. Two examples of arrays with localized edge modes are shown in Fig.~\ref{figS1} for radii of $210$~$\mu\textrm{m}$ [Fig.~\ref{figS1}(a)] and $340$~$\mu\textrm{m}$ [Fig.~\ref{figS1}(b)], with dashed black lines indicating the truncating circles. The propagation constants, sorted in descending order, reveal the formation of quasi-bands. One such quasi-band contains modes localized at the array edges. For $r=210$~$\mu\textrm{m}$ (with $N=76$ waveguides), five edge modes were identified with indices $n=31$, $32$, $33$, $35$, $36$. These edge modes arise due to the flatness of their angular Bloch band, as will be demonstrated by an analytical model below. Fig.~\ref{figS1}(a) shows four examples of such edge modes, alongside a delocalized bulk mode with index $n=44$. Similar to the case with smaller $d$, a slight change in the truncation radius, leading to addition or removal of a layer of waveguides, results in disappearance of these edge modes. For a larger radius of $340$~$\mu\textrm{m}$ (with $N=191$ waveguides), we also found edge-localized modes with indices $n=91$, $92$, $93$, $94$, $95$, as shown in Fig.~\ref{figS1}(b). The intensity profiles of four edge modes are presented in the figure alongside a delocalized mode with index $n=68$. Thus, the existence of edge states is scale-invariant and is primarily determined by the truncation radius and the internal structure (quasi-periodicity) of the array, which enables the formation of flat angular Bloch bands.

}}


\section{Modal content for single-site excitations}\label{sec3}

In the experiments with truncated quasiperiodic arrays, we use single-site excitations. In this setup, a good localization of the linear edge eigenmodes ensures that the excitation of a single waveguide has the largest dominant overlap with the corresponding edge modes. These modes are excited with the largest weights and dominate the propagation dynamics in the linear case. Since our lattice does not change with distance $z$, the linear eigenmodes do not exchange energy during propagation. The modal content for different excitations is depicted in Fig.~\ref{figS2} for $d=2.3$, where we show the weights $|c_n|^2$ of the excited eigenmodes with $c_n=\int w_n^*(\br) \psi(\br,z=0) d^2\br$ (excited waveguide is marked with highlighted circle). Only in the case of the outermost waveguide excitation (red circle) for a truncation radius of $r=260$~$\mu\textrm{m}$ the modes are predominantly localized near the edge with the largest weights (see red dot with $|c_{85}|^2\approx0.57$). All other eigenmodes remain weakly excited and do not contribute significantly to the dynamics, producing only a small background at the noise level. In contrast, a single-site beam launched into other edge waveguides for this radius or into any edge waveguides for truncation radii of $250$ and $270$~$\mu\textrm{m}$ leads to the simultaneous excitation of many modes. Most of these modes are delocalized and extend over the entire lattice, resulting in pronounced diffraction over the sample length. This analytical treatment clearly explains the dynamic properties of the light observed in the experiment for different single-site excitations and truncation radii.

\begin{figure*}[t]
\centering
\includegraphics[width=0.85\linewidth]{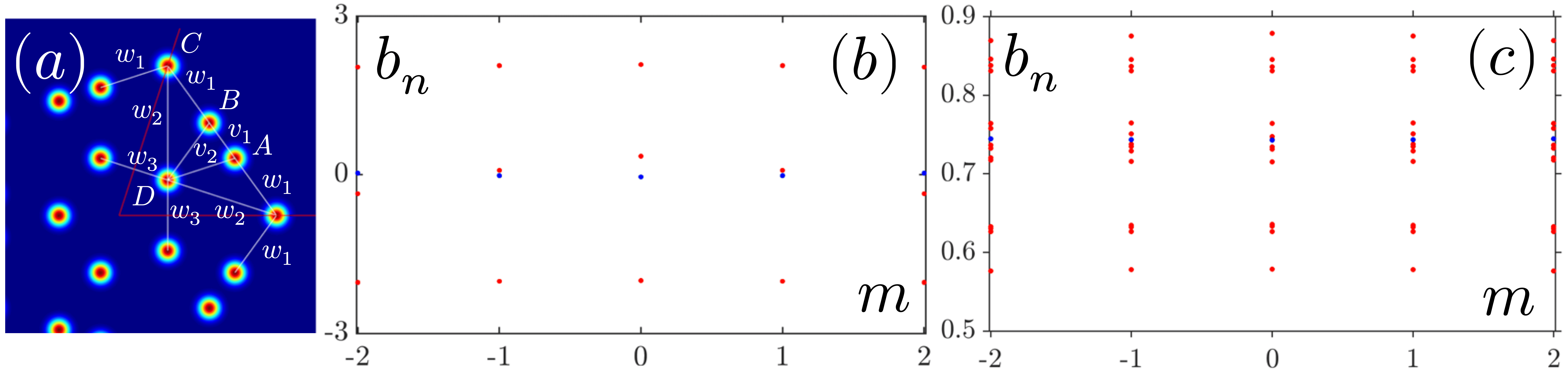}
\caption{
(a) Sketch of the pentagon Penrose tiling (P1) waveguide array with the coupling coefficients $v_i$ and $w_i$ used in the analytical model and the red lines indicating the angular unit cell. (b) Propagation constants $b$ of the analytical model as a function of the orbital pseudo-momentum $m$ for $v_1 = 2$, $v_2 = 0.2$, $w_1 = 0.2$, $w_2 = 0.01$ and $w_3 = 0.2$. (c) Numerically calculated dependence of $b$ on $m$ for a Penrose array quasicrystal with a break-off radius of $r=160$~$\mu\textrm{m}$. The blue dots correspond to the edge modes.
}
\label{figS3}
\end{figure*}


\section{Rotational properties of a quasicrystal waveguide array}\label{sec4}
\subsection{Orbital angular pseudo-momentum}

The quasicrystal waveguide array differs from a standard photonic crystal waveguide in the nature of its bulk symmetries. While the infinite photonic crystal constitutes a bulk owning both discrete translational symmetry and rotational point symmetry around some axes, in the case of a bulk based on a Penrose tiling structure, only the discrete rotational symmetry about certain points remains valid. Since the discrete translational symmetry is lost, Bloch's theorem is no longer applicable and the standard concepts of linear Bloch modes and Bloch bands are no longer suitable analytical tools. However, the correct choice of the axis of rotation in the bulk or in a finite structure based on it, as chosen in the present work, ensures that the discrete rotational symmetry of order $N_r=5$ is a good symmetry of the linear Hamiltonian in Eq.~(1). Mathematically speaking, this means that the Hamiltonian commutes with the $5$ elements of the point group $C_{5}$, which is formed by discrete rotations by a $2\pi/5$ angle. Thus, in this sense, it was proved that the eigenmodes of such systems can be interpreted as \emph{angular Bloch modes}, fulfilling an angular
version of Bloch theorem~\cite{Ferrando2005c} and representing an angular version of it. The eigenvalues of the elementary rotation operator $\mathcal{C}_{5}$ are the roots of unity of order $N_r=5$: $\mathcal{C}_{5}\psi(r,\theta)=\psi(r,\theta+2\pi/5)=e^{i2\pi m/5}\psi(r,\theta)$, with $m=0$, $m=\pm1$ and $m=\pm2$. Therefore, we can classify the eigenmodes of the quasicrystal waveguide by using the integer $m$ as a suitable mode designation. In the case of continuous rotational symmetry ($N_r\rightarrow\text{\ensuremath{\infty}}$), $m$ becomes the usual eigenvalue of the 3rd component of the orbital angular \emph{momentum} (OAM), which is usually represented by $l$. For this reason, we refer to $m$ as orbital \emph{pseudo-momentum} (OAPM), following the terminology of periodic systems. In the language of group theory, each value of $m$ denotes each of the 5 irreducible representations of the group $C_{5}$ in which all modes must be grouped~\cite{hamermesh64}. Since the linear Hamiltonian in Eq.~(1) is real, the system also has time reversal symmetry $T$ ($z\rightarrow-z$ invariance in the case of the waveguide), so that $m=\pm1$ and $m=\pm2$ modes appear as degenerate doublets, while $m=0$ modes are singlets~\cite{hamermesh64}.

\subsection{Angular Bloch bands}

The existence of an angular version of Bloch's theorem allows us to introduce the concept of Bloch's angular bands. For each value of the OAPM $m$, we find a number of modes with different values of $b$, labeled with an angular band index $a$ such that $b=b(m,a)$. In this way, all modes can be classified in terms of angular bands, just as ordinary Bloch modes with linear Bloch momentum $k$ in linear periodic systems are grouped into bands~\cite{Ferrando2005c}. Bloch angular bands are a useful tool for analyzing the existence of edge states in quasicrystalline waveguides.

We have seen that at very specific radii, where light localization occurs at the edges for certain eigenmodes, these modes appear as almost degenerate. The fact that these modes appear as nearly degenerate and that their degeneracy is $5$ can be explained by two simultaneous factors: on the one hand, the strong localization of the modes and, on the other hand, their classification with respect to the $5$ irreducible representations of the $C_{5}$ rotational symmetry group of the array. The strong localization of these modes indicates that the coupling of light between adjacent sites is relatively all, so that their $b$ value corresponds to that of the fundamental mode of a single isolated waveguide. Moreover, all modes are Bloch angular modes~\cite{Ferrando2005c} due to the fivefold rotational invariance of the array. They are grouped in angular Bloch bands $b(m,a)$, which are given by the angular pseudo-momentum $m$ and the band index $a$. The angular Bloch bands for strongly localized modes are special, because their coupling with the neighbors in adjacent angular unit cells (each angular sector acts as an angular unit cell) is negligible for all values of $m$. Therefore, they cluster into almost flat bands that merge into the $b$ value of the fundamental mode of a single waveguide. Although it is possible to find strongly localized states both in the bulk and at the edge, the localized modes at the edge are the states with the strongest localization. This is due to the particular geometry of the edge in the truncated waveguide at certain radii. Accordingly, strongly localized modes at the edges form the flattest angular Bloch band, which consists of $5$ states corresponding to linear combinations of the $m=0,$ $\pm1$, $\pm2$ almost degenerate states of the irreducible representation of the $C_{5}$ rotation group. These $5$ modes have a value of the propagation constant that is almost equal to that of the fundamental mode of a single waveguide.

\subsection{Simple model with localization at edges}

To illustrate the above properties more clearly, let us consider a simple modified discrete version of a quasicrystal waveguide array, which has the geometry of a pentagonal Penrose tiling (P1) with only $20$ sites [see Fig.~\ref{figS3}(a), which illustrates a single angular unit cell, highlighted by red lines]. The pentagon is divided into $5$ angular sectors with $4$ sites each [see Fig.~\ref{figS3}(a)]. Each angular sector acts as an angular unit cell. The unit cell is formed by sites C and D at the vertex of the outer and inner rings, respectively, and sites A and B, which lie symmetrically on the side of the outer ring. Using angular Bloch's theorem, the eigenstates of the system are described by a four-dimensional vector of form\@
\[
\Psi_{m}=e^{i2\pi m/5}u_{m}=e^{i2\pi m/5}\left(\begin{array}{c}
c_{m,A}\\
c_{m,B}\\
c_{m,C}\\
c_{m,D}
\end{array}\right),\,\,\,\,m=0,\pm1,\pm2,
\]
where the components of the complex $4$-vector are the amplitudes at every site. Site C is coupled to sites A and B with the coefficient $w_{1}$ and to site D with the coefficient $w_{2}$. Site D is coupled to sites A and B with the coefficient $v_{2}$, to site C with the coefficient $w_{2}$, and with the neighboring sites D with the coefficient $w_{3}$. Finally, sites A and B are coupled with the coefficient $v_{1}$ and with site D with the coefficient $v_{2}$. This simple coupling arrangement allows us to write the reduced Hamiltonian associated with the $m$ Bloch sector, whose eigenvectors are the $u_{m}$ angular Bloch functions of the $m$ sector and whose eigenvalues are their corresponding $b(m,a)$ propagation constants ($a=1,2,3,4$):
\begin{align*}
& \left[\begin{array}{cccc}
0 & v_{1} & w_{1}e^{-i\frac{2\pi}{5}m} & v_{2}\\
v_{1} & 0 & w_{1} & v_{2}\\
w_{1}e^{i\frac{2\pi}{5}m} & w_{1} & 0 & w_{2}\left(1+e^{-i\frac{2\pi}{5}m}\right)\\
v_{2} & v_{2} & w_{2}\left(1+e^{i\frac{2\pi}{5}m}\right) & w_{3}\left(e^{i\frac{2\pi}{5}m}+e^{-i\frac{2\pi}{5}m}\right)
\end{array}\right]
\end{align*}
The eigenvalues of $b(m,a)$ provide the $4$ Bloch angular bands (each for each value of $a$). In Fig.~\ref{figS3}(b) we give the Bloch angular bands for a particular choice of couplings where we have $w_{1}$ and $w_{2}$ much smaller than the other coupling coefficients.
This has the remarkable effect that we obtain an almost flat band with $b=0$ [blue dots in Fig.~\ref{figS3}(b)]. In this case, this value corresponds to the propagation constant of the fundamental mode of an isolated waveguide. This fact can be understood by noticing that in the absence of all couplings (isolated waveguide) the eigenvalues are trivially zero. To understand the reason for the appearance of this flat band, it is interesting to analyze the so-called dimerized limit, as in the Su-Shrieffer-Heeger (SSH) model for condensed matter
\cite{Asboth2016}. We set $w_{1}=w_{2}=v_{2}=0$ to completely decouple (dimerize) the
C-sides from the remainder. In this way, we can determine eigenvectors and eigenvalues analytically. We obtain two symmetrical flat bands with
$b=\pm\left|v_{1}\right|\ne0$, one angular dispersive non-flat band with $b_{m}=2w_{3}\cos\left(2\pi m/5\right)$ and, interestingly, a completely flat band with $b=0$. The eigenvector associated with the latter flat band is identical for all values of $m$ and is given by
\[
u_{m}=\left(\begin{array}{c}
0\\
0\\
1\\
0
\end{array}\right),\,\,\,m=0,\pm1,\pm2,
\]
thus indicating full localization at the vertexes of the waveguide
edge located at C sites. As expected, a hallmark of strong localization
at the edge is the appearance of angular flat bands with a value of
$b$ approximately equal to that of the fundamental mode of an isolated
waveguide.

This behavior is also observed in our simulations of realistic quasicrystal
waveguide arrays. In Fig.~\ref{figS3}(c) we represent the angular Bloch bands
associated to the modal spectrum for the truncation radius $r=160$~$\mu\textrm{m}$,
as in Fig~2(a) of the main text of the Letter, but with equal widths $w_x=w_y=5.0$~$\mu\textrm{m}$. In this plot, we choose eigenmodes which are simultaneously
eigenstates of the Hamiltonian and of the discrete rotation operator
and thus they can be classified according to their $m$ value. As
in our simplified model, we recognize (blue dots) an almost flat band
at the $b$ value of the fundamental mode of an isolated waveguide,
although in our array, waveguides, where the edge modes are predominantly localized, are not isolated.
These $5$ eigenmodes correspond to the strongly localized modes at the
edge of the quasicrystal array.


\begin{figure}[t]
\centering
\includegraphics[width=0.9\linewidth]{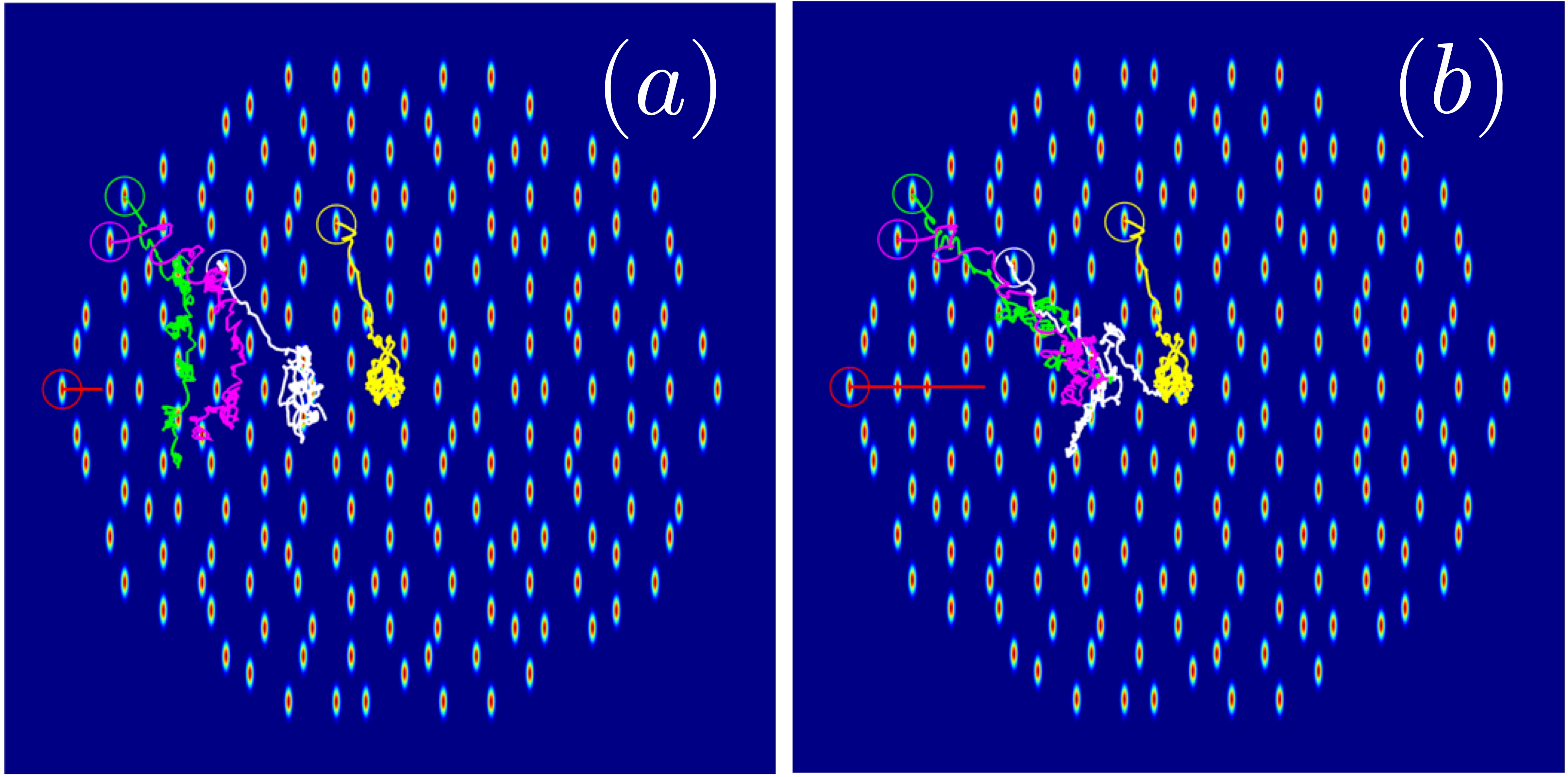}
\caption{{{Trajectories of the integral beam center superimposed on the array profile for single-site excitations of different waveguides (indicated by circles) at two power levels: $U=0.01$~(a) and $U=0.25$~(b). The color of each trajectory corresponds to the color of circle around respective excited waveguide, providing a clear visual match between the excitation position and subsequent trajectory of its integral center. The array is shown within the window $x,y\in[-300\,\mu \textrm{m},+300\,\mu \textrm{m}]$.}}
}
\label{figS4}
\end{figure}

{{

\section{Trajectories of the integral beam center for excitation of different waveguides}\label{sec5}

To further highlight the differences in localization properties observed upon excitation of different waveguides of the array, Fig.~\ref{figS4} presents the trajectory of the integral center of the beam calculated using the expressions
\[
\langle x(z) \rangle=U^{-1} \int x|\psi|^2 d^2\br, \qquad \langle y(z) \rangle=U^{-1} \int y|\psi|^2 d^2\br,
\]
where $U=\int |\psi|^2 d^2\br$, and superimposed on the array profile. The trajectories are shown as curves of different colors, with each curve's color corresponding to the color of circle indicating the initially excited waveguide. For comparison, Fig.~\ref{figS4}(a) illustrates the result of linear propagation ($U=0.01$), while Fig.~\ref{figS4}(b) shows propagation in weakly nonlinear case ($U=0.25$). For these simulations the propagation distance was set to $500$ sample lengths, or $5000\,\textrm{cm}$, emphasizing that the observed effect persists over distances much larger than the experimental sample length. In Fig.~\ref{figS4}(a), it is evident that only the excitation of the edge waveguide, marked by the red circle, remains well localized. Slight deviation of the red trajectory from the center of the excited edge waveguide is due to small radiation at the initial stages of propagation. In all other cases illustrated in this figure, the trajectories of the integral beam center are significantly elongated, they indicate on the displacement of the integral center towards the center of the array that is accompanied by complete delocalization of the beam, with field spreading across the entire array. Notably, exciting waveguides other than the one shown with the red circle, and positioned along the array's mirror symmetry axes both at the edge and in the bulk of the array leads to beam broadening (the only exception is the excitation of the waveguides analogous to red one, but corresponding to rotation of the structure by an angle $2\pi n/5$, as a consequence of $\mathcal{C}_5$ symmetry of the array). Similar results were obtained for other excitation positions (not shown in the figure). In weakly nonlinear regime [Fig.~\ref{figS4}(b)], the red curve lengthens, indicating on nonlinearity-induced coupling with bulk modes (similar effect is observed for exact soliton bifurcating from linear mode centered in this waveguide, as discussed in the main text). As power increases even further, localization strengthens for all excited waveguides, leading eventually to a sharp reduction in the lengths of beam center trajectories and concentration of light in excited channels. This result underscores the significance of the truncation radius and highlights the varying localization effects when different waveguides are excited within the quasicrystal array.

}}

\begin{figure}[t]
\centering
\includegraphics[width=1\linewidth]{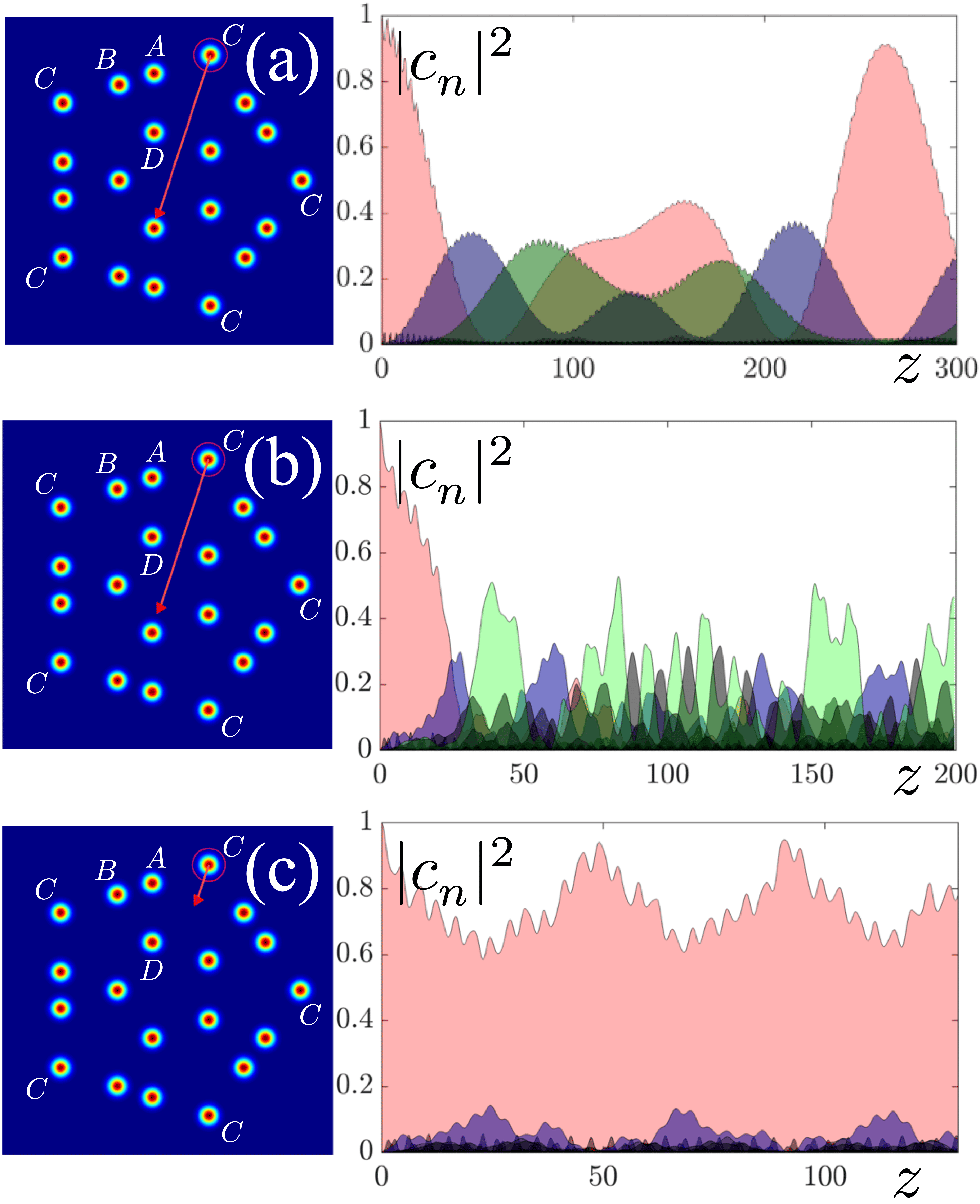}
\caption{
{{
The trajectories (left column) of the pulse center of mass overlayed on a sketch of the waveguide array profile, which serves as the analytical model to explain the localization mechanism. The excited waveguides C are highlighted with a circle, and the triangle marks the farthest point reached by the trajectory from the excited waveguide. The site fractions $|c_n|^2$ dependence on the waveguide states along the distance $z$ is shown in the right column, with the most involved states highlighted in red, blue, and green. Panels illustrate (a) the linear regime, (b) the nonlinear regime below the localization threshold, and (c) the nonlinear regime above the localization threshold.
}}
}
\label{figS5}
\end{figure}

{{

\section{Nonlinear localization as a critical phenomenon in a simplified model}\label{sec6}

In order to clarify the aspects of localization at edges, we resort
to the same simplified $20$ sites pentagon Penrose tiling (P1) discrete model used in 
section~\ref{sec4}. We consider now the propagation in $z$ of the $n=20$ components
$c_{n}(z)=c_{j\alpha}(z)$ of the field amplitudes at the site located
at the angular section $j$ (ranging from $1$ to $5$) and at position
$\alpha=A$, $B$, $C$, $D$ at every site, as depicted in Fig.~\ref{figS3}(a). The nonlinear
propagation equations for them are:
\[
i\frac{dc_{j\alpha}}{dz}=g\left|c_{j\alpha}\right|^{2}c_{j\alpha}+\sum_{\left\langle \alpha j,\alpha'j'\right\rangle }v_{j\alpha,j'\alpha'}c_{j'\alpha'},
\]
where $g$ is the local nonlinear coefficient and $v_{j\alpha,j'\alpha'}$
are the coupling coefficients among sites, which in this case we consider
non-vanishing only for near neighbor sites. In this way, the only
non-zero coefficients are those appearing in Fig.~\ref{figS5}(a), as we considered
in a previous section.

We analyze the propagation of light injected in a single-edge site
C in the first angular section $j=1$ for different values of the
nonlinear coefficient $g$. First, we consider that the initial state
is normalized to $1$, so that $c_{1C}(0)=1$ whereas the rest of the
initial state components vanish. We present in Fig.~\ref{figS5} different
examples of propagation of this edge state for diverse values of $g$.
We note that distances $AC$, $CD$, and $DD'$ (distance between two
neighboring D sites) are the same and, at the same time, distance
$AB$ is the smallest whereas distance $CD$ is the largest. Accordingly,
$v_{1}>w_{1}=w_{3}=v_{2}>w_{2}$ and we choose for this analysis $v_{1}=2,\,w_{1}=0.2$
and $w_{2}=0.01$. In the right column, we show the evolution in $z$
of the fractions of light intensity at every site, i.e., the modulus
square of the components of the field amplitude $\left|c_{n}\left(z\right)\right|^{2}$.
In the left column, we represent the trajectory of the center of mass
for our particular initial condition of illuminating site $1C$ only.
We define the center of mass evolution in the discrete case simply
as the weighted average $\left\langle \mathbf{r}\left(z\right)\right\rangle =\sum_{n}\mathbf{r}_{n}\left|c_{n}\left(z\right)\right|^{2}$. 

In Fig.~\ref{figS5}(a) we present the linear propagation ($g=0$) of the
center of mass, in which, firstly, we observe that its trajectory
is linear and, secondly, that it moves well toward the center penetrating
deeply inside the ``bulk'' part of the array. The linear behavior
of the trajectory is due to the mirror symmetry of the array with
respect to the axis $r_{CO}$ passing through C and the center of
the pentagon $O$, which forces all components of mirror symmetric
sites with respect to this axis to be equally excited during propagation.
This can be confirmed by checking the evolution of $\left|c_{n}\left(z\right)\right|^{2}$
at different sites [right column Fig.~\ref{figS5}(a)]. The linear evolution
is dominated by light at C sites (red, green, and blue curves).
Each curve is doubly degenerated corresponding to the two symmetric
C sites with respect the mirror axis $r_{CO}$, which forces the center
of mass to propagate along a straight line.
Consequently, in this case, light oscillates between the C waveguides, with minimal penetration into the bulk of the array---unlike in the primary array of waveguides used in the experiment.

The deep penetration length and the straight line trajectory of the
center of mass features hold also in the nonlinear case ($g\ne0$).
However, the penetration length experiments an abrupt qualitative
change of behavior at a critical value of the coupling constant, which,
for our particular choice of linear coupling coefficients occurs at
$g_{c}\approx0.76$. In Fig.~\ref{figS5}(b) we show the evolution of the
center of mass and of the site fractions $\left|c_{n}\left(z\right)\right|^{2}$
just below the critical point $g=0.75<g_{c}\approx0.76$. We still
see a high value of the penetration length inside the ``bulk''.
However, the analysis of the fraction evolution indicates that now
the dominating modes correspond mostly to D sites combined with C
sites. The critical nature of the mechanism fully unveils when we
increase the nonlinear coefficient just a small quantity above $g_{c}$,
as one can appreciate in Fig.~\ref{figS5}(c). For $g=0.77>g_{c}\approx0.76$,
the penetration length collapses to a value considerably smaller than
the distance from the site C to the inner pentagon thus indicating
critical localization at the $1C$ edge, which corresponds to the originally
illuminated waveguide. The evolution of the site fractions $\left|c_{n}\left(z\right)\right|^{2}$
also reflects, in turn, this critical behavior by pointing out a sudden
change in the dominant modes, which now correspond to the single $c_{1C}$
component of the original $1C$ site combined with small neighboring
D sites components.

We can go a step further in order to provide a better understanding
of this critical mechanism by resorting to an even simpler model,
which nevertheless still grasp \emph{qualitatively} the fundamental
features of the phenomenon near the critical point. As mentioned above,
near the critical point only C sites (mostly the initial $1C$ site)
and $D$ sites (mostly the two nearest two D sites to the $1C$ site) are
dominant. Since by symmetry the two D sites ($1D$ and $5D$) are always
equally excited in evolution (they appear always degenerate) we can
consider that they always form a single symmetric $DD$ dimer state.
Thus, we can define a simplified two-mode version of light propagation
involving the $1C$ site and the $DD$ dimer components $c_{1C}$ and $c_{DD}$. We expect this nonlinear model to be approximately valid only near
$g_{c}$. Its evolution equations are
\begin{align}
i\frac{dc_{1C}}{dz} & =g\left|c_{1C}\right|^{2}c_{1C}+w\,c_{DD}\nonumber \\
i\frac{dc_{DD}}{dz} & =g\left|c_{DD}\right|^{2}c_{DD}+w\,c_{1C},\label{eq:two_mode_NL_eqs}
\end{align}
where $w$ is the effective edge-dimer coupling constant, which is
in principle different from $w_{2}$. We now use a standard equivalence
\cite{Ferrando2017}, which permits to transform the previous propagation
equations into a nonlinear evolution equation for a spin in an effective
magnetic field $\mathbf{\Omega}$, using the following equivalences:
\begin{equation}
\left|C\right\rangle =\left(\begin{array}{c}
c_{1C}\\
c_{DD}
\end{array}\right)\Longleftrightarrow\mathbf{S}=\left(i\begin{array}{c}
c_{1C}^{*}c_{DD}+c_{DD}^{*}c_{1C}\\
\left(c_{1C}^{*}c_{DD}-c_{DD}^{*}c_{1C}\right)\\
\left|c_{1C}\right|^{2}-\left|c_{DD}\right|^{2}
\end{array}\right)\label{eq:S_vector}
\end{equation}
and
\[
\mathbf{\Omega}=\left(\begin{array}{c}
2w/g\\
0\\
\left|c_{1C}\right|^{2}-\left|c_{DD}\right|^{2}
\end{array}\right)
\]
 together with the equation of motion $d\mathbf{S}/dt=\mathbf{\Omega}\times\mathbf{S}$.
The Hamiltonian of the spin model is simply $H=\mathbf{\Omega}\cdot\mathbf{S}$,
which, in turn, is a constant of motion. Therefore
\[
H=\mathbf{\Omega}\cdot\mathbf{S}=2wS_{1}+S_{3}^{2}=H(0)=1
\]
since according to the initial condition $S_{1}\left(0\right)=0$
and $S_{3}\left(0\right)=1$. Taking into account that the equivalent
spin is normalized to $1$, we arrive at the simple condition 
\begin{equation}
S_{2}^{2}-\left(1-S_{3}^{2}\right)\left(1-\frac{g^{2}}{4w^{2}}\left(1-S_{3}^{2}\right)\right)=0.\label{eq:condition_for_S2_S3}
\end{equation}

In this case, the center of mass position is given by $\left\langle \mathbf{r}\left(z\right)\right\rangle =\mathbf{r}_{1C}\left|c_{1C}\left(z\right)\right|^{2}+\mathbf{r}_{DD}\left|c_{DD}\left(z\right)\right|^{2}$,
which in terms of $S_{3}$ reads
\[
\left\langle \mathbf{r}\left(z\right)\right\rangle =\frac{\mathbf{r}_{1C}+\mathbf{r}_{DD}}{2}+\frac{\mathbf{r}_{1C}-\mathbf{r}_{DD}}{2}S_{3}(z).
\]
This expression shows that the center of mass follows a straight line
trajectory along the symmetry axis $r_{CO}$ from the $1C$ site ($\left\langle \mathbf{r}\left(z\right)\right\rangle =\mathbf{r}_{1C}$
when $S_{3}=1$) to the dimer center ($\left\langle \mathbf{r}\left(z\right)\right\rangle =\mathbf{r}_{DD}$
when $S_{3}=-1$) in which $S_{3}$ acts as the straight line parameter.
If we use this symmetry axis as a new coordinate axis, we set the
origin at the midpoint $\mathbf{r}_{M}=\left(\mathbf{r}_{1C}+\mathbf{r}_{DD}\right)/2$
and take into account that $\mathbf{r}_{1C}-\mathbf{r}_{DD}$ is a
vector pointing from the dimer center to $1C$ along $r_{CO}$, we can
express the center of mass position as
\begin{equation}
\left\langle \mathbf{r}\left(z\right)\right\rangle -\mathbf{r}_{M}=\frac{d}{2}S_{3}\left(z\right)\hat{\textbf{n}}=s\left(z\right)\hat{\textbf{n}},\label{eq:def_center_of_mass_S3}
\end{equation}
where $\mathbf{\hat{n}}$ is the unit vector along $r_{CO}$, $d$
is the distance between the site $1C$ and the dimer center and $s$
is the new center of mass coordinate. By setting $d=2$, we see we
can use $S_{3}$ directly as the center of mass coordinate. On the
other hand, by using the propagation equations, the definition of
the center of mass in terms of the site fractions and the form of
the second component of $\mathbf{S}$ in (\ref{eq:S_vector}), we
can prove that 
\begin{equation}
\frac{d\left\langle \mathbf{r}\left(z\right)\right\rangle }{dz}=\frac{ds\left(z\right)}{dz}\mathbf{\hat{n}}=2\frac{w}{g}dS_{2}\left(z\right)\mathbf{\hat{n}}.\label{eq:def_velocity_center_of_mass_S2}
\end{equation}
Therefore, the constant of motion condition for the equivalent spin
model (\ref{eq:condition_for_S2_S3}) can be rewritten using equations
(\ref{eq:def_center_of_mass_S3}) and (\ref{eq:def_velocity_center_of_mass_S2})
in the appealing form of a conservation law for the mechanical energy
of a particle in a $1D$ potential
\begin{equation}
\frac{1}{2}\left(\frac{ds}{dz}\right)^{2}+V\left(s\right)=0,\label{eq:mechanical_energy}
\end{equation}
where the potential is ($d=2$)
\begin{equation}
V\left(s\right)=-\frac{1}{2}\left(1-s^{2}\right)\left(4\frac{w^{2}}{g^{2}}-1+s^{2}\right).\label{eq:potential}
\end{equation}

We see that the potential has a non-analytical dependence on the coupling
constant $g$, a signature typical of critical behavior. The nature
of $V$ determines the type of trajectory followed by the particle
initially at site $1C$ when $s(0)=1$. This trajectory is determined
by the turning points given by the condition $V=0$, where ``velocity''
vanishes $ds/dz=0$, according to equation (\ref{eq:mechanical_energy})
For $g<2w$ the trajectory has a single turning point at $s_{t}^{(-)}=-1$.
Physically, the center of mass starts moving from the site $1C$, penetrates
well into the ``bulk'', reaching the inner pentagon at the $DD$ dimer
center, and then bouncing back to the corner site at $1C$. This scenario
is the same as the case of deep penetration into de ``bulk'' when
$g<g_{c}$ analyzed in full simulations. For $g>2w$ and the initial
condition $s(0)=1$, the turning point drastically turns into a finite
value given by $s_{t}^{(+)}=\sqrt{1-4w^{2}/g^{2}}\ge0.$ The penetration
length is now restricted to distances smaller than the one from the
$1C$ site ($s=1$) to the middle point $M$ ($s=0$). The trajectory avoids
deep penetration into the ``bulk'' and provokes a sudden localization
of light around the corner, identically to the $g>g_{c}$ case in
full simulations. However, now we can give explicit analytical expressions
for the critical value and the behavior around it. We use the penetration
length, mathematically described by the turning point $s_{t}$, as
the order parameter. The critical value for $g$ is that for which
$s_{t}^{(+)}$ is no longer a real solution. This clearly happens
when $g=g_{c}=2w$, for which $s_{t}^{(+)}=0$. For $g<2w$ the only
allowed solution is $s_{t}^{(-)}=-1$. We can use the explicit analytical
expression for $s_{t}^{(+)}$ to characterize this critical behavior:
\[
s_{t}=\begin{cases}
s_{t}^{(+)}=\sqrt{1-4\frac{w^{2}}{g^{2}}}\stackrel{g\gtrsim2w}{\approx}\frac{1}{w^{1/2}}\left(g-g_{c}\right)^{1/2}, & g>g_{c}\\
0, & g=g_{c}=2w\\
s_{t}^{(-)}=-1, & g<g_{c}.
\end{cases}
\]
Written in this way, we can interpret edge localization as a critical
phenomenon characterized by a first-order phase transition occurring
at the critical value $g_{c}=2w$ and with order parameter $\beta=1/2$. 

}}

\begin{figure*}[t]
\centering
\includegraphics[width=1\linewidth]{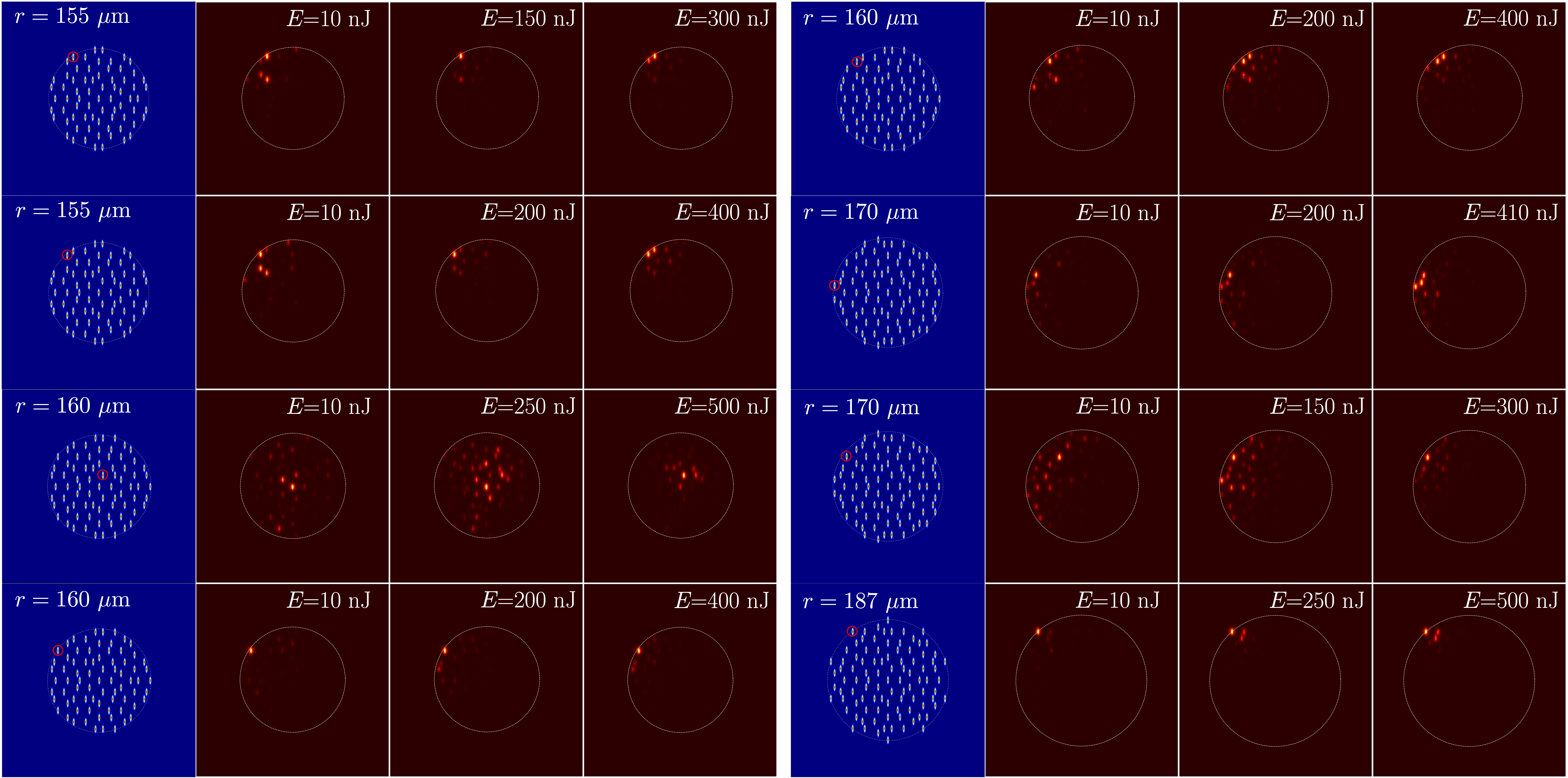}
\caption{
Waveguide array profiles (blue background) and experimental output intensity distributions (maroon background) for different input pulse energies for $r=155$~$\mu\textrm{m}$, $r=160$~$\mu\textrm{m}$, $r=170$~$\mu\textrm{m}$, and $r=187$~$\mu\textrm{m}$. The red circles indicate the excited waveguide}
\label{figS6}
\end{figure*}


\section{Output experimental intensity distributions}\label{sec7}

In the main text of the Letter, we demonstrated that a well-localized edge mode is achieved with a truncation radius of $260$~$\mu\textrm{m}$. Notably, a slight decrease or increase in this radius, such as removing or adding a layer of waveguides, prevents localization at the array's edge. We also found that similar localization in the linear regime occurs for radii of $160$ and $187$~$\mu\textrm{m}$. This is illustrated in Fig.~\ref{figS6}, which presents arrays of waveguides with radii of $r=155$~$\mu\textrm{m}$, $r=160$~$\mu\textrm{m}$, $r=170$~$\mu\textrm{m}$, and $r=187$~$\mu\textrm{m}$. Experimentally, we excited the outer waveguides, indicated by red circles in the panels with array profiles. For a radius of $155$~$\mu\textrm{m}$ and low pulse energies, diffraction into the bulk of the array is observed. At higher energies, light predominantly localizes in two edge waveguides, between which switching occurs along the sample's propagation length. With further increased energy, the light becomes localized solely in the excited waveguide. For a radius of $160$~$\mu\textrm{m}$, illumination of the outermost waveguide leads to the excitation of the edge mode shown in Fig.~2(a) of the main text, and, therefore, the light stays in the excited waveguide. We confirmed that there is no localization when a waveguide in the array's bulk or another waveguide at the edge (except for the outmost waveguides corresponding to the $5$ localized edge modes) is excited. In these scenarios, localization gradually increases with rising input pulse energy. A similar absence of localization in the linear regime is observed for a radius of $170$~$\mu\textrm{m}$. Interestingly, for this radius excitation of the waveguide that is outermost for a radius of $160$~$\mu\textrm{m}$ results in strong diffraction into the array's bulk, despite no structural changes occurring in the close vicinity with this waveguide after the radius shift. This underscores the critical role of the truncation radius in achieving edge localization. For a radius of $187$~$\mu\textrm{m}$, we observe strong localization at the sample's edge in the case of outmost channel excitation. Thus, we have both theoretically and experimentally demonstrated that edge localization occurs only for specific truncation radii for the quasiperiodic structure. Even a small change in this radius that alters structure causes localization to vanish. This light localization facilitates the presence of solitons without a power threshold.


\begin{figure}[t]
\centering
\includegraphics[width=1.0\linewidth]{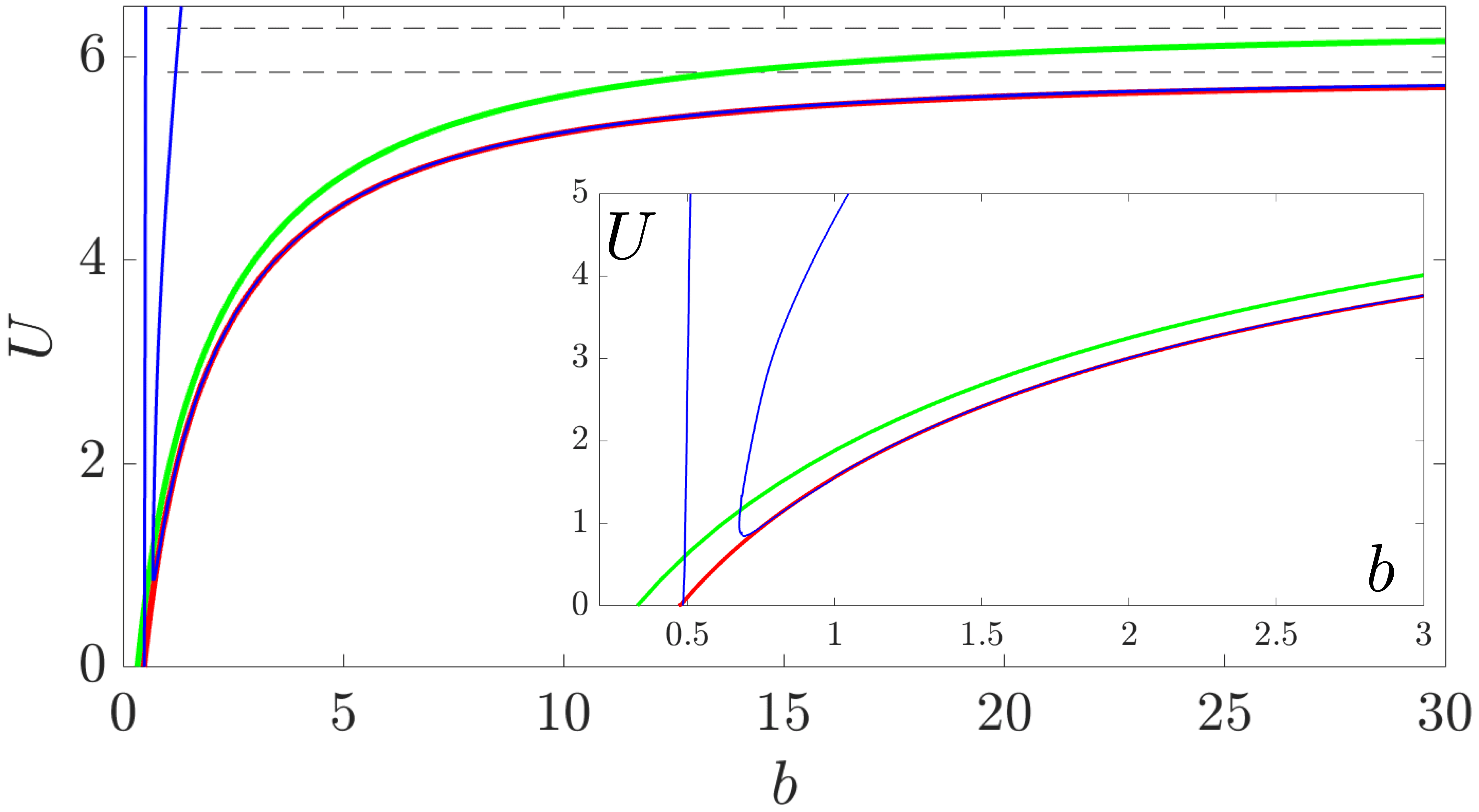}
\caption{
{{
The $U(b)$ dependencies for a single waveguide for the variational approximation (green curve) and numerical calculations (red curve), and two branches of the edge soliton family obtained from numerical simulations for quasicrystal waveguide arrays with $r=260$~$\mu\textrm{m}$ (blue curves). The inset provides an enlarged view of the soliton families near the bifurcation point from linear mode. The horizontal dashed lines represent the asymptotic limits predicted by both the analytical ($U_{cr}=2\pi$) and numerical approaches ($U_{cr}=5.85$).
}}
}
\label{figS7}
\end{figure}

{{

\section{Soliton solutions and wave collapse in a single waveguide}\label{sec8}

In this section, we examine how the presence of a waveguide influences the stability of soliton solutions and the occurrence of wave collapse. It is worth noting that soliton stability and wave collapse have been extensively studied in different areas of physics~\cite{Berge1998, SulemSulem1999, KivsharPelinovsky2000, KartashovMalomed2011, Fibich2015, Malomed2024} and it is now particularly well understood for uniform Kerr media, while analytical results for Kerr media with spatially inhomogeneous refractive index landscapes are scarce. In 2D or 3D systems one of the main obstacles for obtaining analytical stability criteria is non-integrability of corresponding evolution equations and the absence of analytical expressions for soliton solutions. On this reason, stability of 2D solitons in inhomogeneous optical media is most frequently studied using numerical solution of corresponding linearized eigenproblems for perturbations of stationary soliton states (in this case, the absence of growing perturbations at least implies the existence of stable solitons, even though it cannot prove that the collapse is completely arrested by the refractive index modulation). While non-integrability of 2D equation possessing soliton solutions is often seen as a technical obstacle, the approximate analytical methods, such as the variational approximation, can still be employed to construct soliton solutions. Moreover, numerical simulations can effectively handle these equations using modern computational power~\cite{Yang2010}.

It is known that in 2D uniform Kerr medium the collapse occurs for light beams when their power (norm) $U = \int |\psi|^2 d^2\br$ exceeds a certain critical value, $U_{cr}$, which has been numerically determined to be approximately $U_{cr}\approx5.85$. This is exactly the power of Townes soliton. The beam with power below critical one diffract in the course of propagation. Analytically, this critical power can be predicted using the variational approach with a Gaussian ansatz, that yields $U_{cr}=2\pi$. The study of soliton states becomes more complex when considering a material with non-uniform refractive index defining, for example, a periodic 2D waveguide array. Nevertheless, it has been shown that such arrays can suppress collapse leading to stable propagation of fundamental soliton states (see, e.g.,~\cite{AcevesTuritsyn1994, AcevesTuritsyn1995, PelinovskyKevrekidisFrantzeskakis2005, KartashovMalomed2009, ChongPelinovsky2011}). Direct linear stability analysis have also confirmed stability of 2D nonlinear states bifurcating from localized linear modes of aperiodic fractal waveguide arrays \cite{ZhangKartashov2024}. All these results hint that periodic or aperiodic modulations of the refractive index of material do have stabilizing action on 2D states in Kerr medium.

In our scenario, we examine an aperiodic quasicrystal waveguide array excited by sufficiently long pulses ($280$~fs), allowing us to neglect temporal evolution. Additionally, since we operate in the normal dispersion regime, increasing intensity does not lead to catastrophic pulse compression; instead, it contributes to pulse broadening over time. Spatial dynamics, however, presents a greater challenge. As illustrated in Fig.~3 of the main text of the Letter, the complex geometry of the quasicrystal array results in a non-trivial structure of the family of edge solitons. Here, we mostly focus on analyzing the collapse threshold within this structure. Both numerical simulations and experimental data suggest that wave collapse occurs only for powers significantly higher than those required for light to localize within a single waveguide ($U\approx1$). Thus, spatial collapse can potentially occur only when light is strongly confined within an individual site, effectively isolating it from interactions with neighboring waveguides. Therefore, to analyze wave collapse in our system, we simplify the problem to a single waveguide case, where $\mathcal{R}(\br) = p e^{-\left[x^2/w_x^2+y^2/w_y^2\right]}$. In this case, we can apply the variational approximation, adopting the Gaussian ansatz
\[
\psi(\br,z) = A e^{-\left[x^2/a_x^2+y^2/a_y^2\right]}e^{ibz}
\]
Using the Lagrangian formalism, we first derive the Lagrangian density in our system
\[
\mathcal{L} = \frac{i}{2}\left(\psi^*\psi_z-\psi\psi^*_z\right) - \frac{1}{2}|\nabla \psi|^2+\frac{1}{2}|\psi|^4+\mathcal{R}|\psi|^2.
\]
By substituting the above ansatz into the system's Lagrangian $L=\int \mathcal{L} d^2\br$, and integrating over the transverse space, we obtain the ``averaged'' Lagrangian
\begin{align*}
L=\frac{\pi}{4}A^2a_x a_y &\left( A^2 - 4b - \frac{a_x^2+a_y^2}{a_x^2a_y^2}   \right.\\
&+\left. 
\frac{4pw_xw_y}{\sqrt{(a_x^2+w_x^2)(a_y^2+w_y^2)}}\right).
\end{align*}
The corresponding equations for soliton solution can then be derived from the Euler-Lagrange equations, $\partial L/\partial a_x=\partial L/\partial a_y=\partial L/\partial A=0$, and $U=\pi A^2a_xa_y$, leading to the final results
\begin{align*}
U=&2\pi\left( \frac{a_x}{a_y} -\frac{2pa_xa_y^3w_xw_y}{(a_y^2+w_y^2)\sqrt{(a_x^2+w_x^2)(a_y^2+w_y^2)}} \right),\\
b=&\frac{3}{4 a_x^2} - \frac{1}{4 a_y^2} - \frac{
 p w_x w_y(a_x^2-w_x^2)}{(a_x^2 + w_x^2) \sqrt{(a_x^2 + w_x^2)(a_y^2 + w_y^2)}} = \\
 -&\frac{1}{4 a_x^2} + \frac{3}{4 a_y^2} - \frac{p w_x w_y (a_y^2-w_y^2)}{(a_y^2 + w_y^2)\sqrt{(a_x^2 + w_x^2)(a_y^2 + w_y^2)}}.
\end{align*}
The nonlinear family $U(b)$, obtained by solving these three equations, is shown by the green curve in Fig.~\ref{figS7}. The waveguide parameters used here are identical to those in the main text of the Letter. As one can see, the power $U$ is a monotonically increasing function of the propagation constant $b$, which means that this family of solutions always satisfies the Vakhitov-Kolokolov stability criterion ($d U/d b>0$) for all values of the propagation constant $b$~\cite{VakhitovKolokolov-73}. For comparison, the figure also shows the numerically calculated $U(b)$ dependence for a single waveguide, represented by the red curve. While there is a slight shift between these two curves, their overall behavior is consistent in numerics and variational approaches. It is clear from the figure and the equations that the power is limited from above by the value of $2\pi$, while the numerical result estimates the critical power to be approximately $5.85$, which coincides with the known critical power (universal norm of 2D soliton) for uniform medium. These asymptotic limits are depicted by the horizontal dashed lines. Additionally, Fig.~\ref{figS7} includes two branches (blue curves) of the numerically obtained edge soliton family for the quasicrystal waveguide array with a truncation radius of $r=260$~$\mu\textrm{m}$: The left of these curves bifurcates from the linear edge mode, while the lower part of the right curve practically coincides with $U(b)$ dependence obtained for single waveguide at sufficiently high powers indicating that the analytical approach remains valid for sufficiently large $b$. The complete nonlinear family for the quasiperiodic array is depicted in Fig.~3 of the main text. Furthermore, the bifurcation point of the blue curve coincides with that of the red curve, as shown in the inset, indicating that the propagation constant of the linear edge-localized states matches that of the fundamental mode of a single waveguide. As $b$ increases, the power $U$ of the left blue branch rises sharply due to interactions with nearby linear bulk modes in the spectrum. An analytical approximation for this segment of the family is provided in the next section.

In the case of a radially symmetric waveguide, where $w=w_x=w_y$, this system of three equations simplifies to the following form:
\begin{align*}
U=&2\pi\left(1-\frac{2pw^2a^4}{(a^2+w^2)^2}\right),\\
b=&\frac{1}{2a^2}-\frac{p w^2 (a^2 - w^2)}{(a^2 + w^2)^2}
\end{align*}
with $a=a_x=a_y$. Note that, if $p=0$ (i.e., no waveguide), the result yields $U=2\pi$ and $b=1/(2a^2)$. In this context, this corresponds to the well-known variational prediction for the critical power for the 2D nonlinear Schr\"{o}dinger equation.

Thus, we have derived analytical equations describing the family of self-sustained states for a single waveguide, which can also be applied to describe the nonlinear families of edge solitons in a quasicrystal system at sufficiently high power, where the light is well localized within an individual waveguide. Since these solitons bifurcating from the linear mode of quasicrystal structure also represent simple, non-excited nonlinear states of the system, their stability properties are also usually correctly predicted by the Vakhitov-Kolokolov stability criterion $dU/db>0$ that implies that collapse cannot develop for such stable nonlinear states. If the input beam overlaps efficiently with corresponding solitons, then most of the beam power concentrates in soliton state that then exhibits stable propagation (i.e. it does not collapse). Namely such solitons with power well below critical one were excited in experiments to avoid optical damage of the material. If however, the input beam carries power exceeding critical one, it may collapse despite the presence of the waveguide or array. This was further validated by numerical simulations of nonlinear propagation for different initial beam power, which showed reasonable agreement with these analytical predictions.

}}


\begin{figure}[t]
\centering
\includegraphics[width=1.0\linewidth]{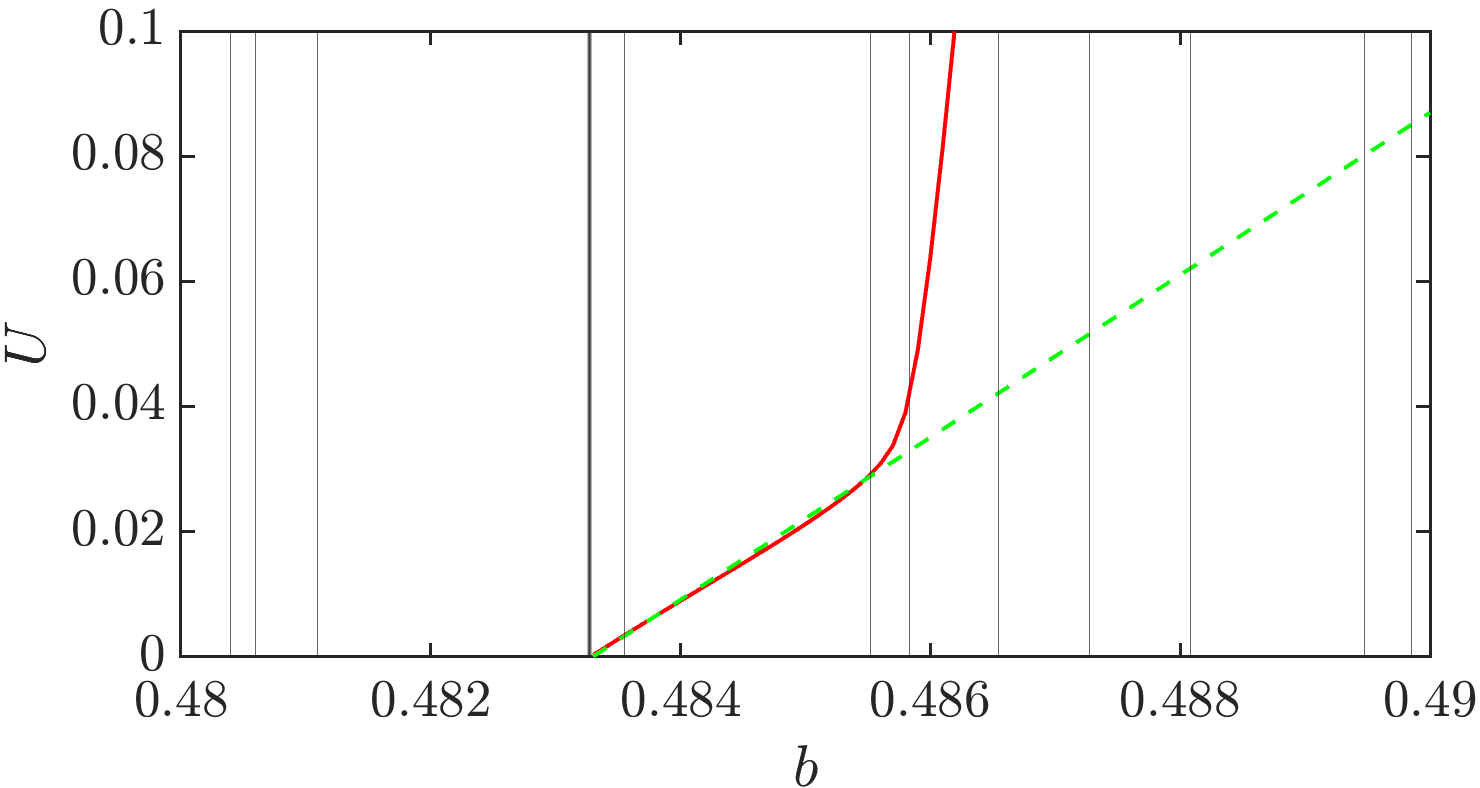}
\caption{
{{
The $U(b)$ dependencies for the edge soliton near the bifurcation point are illustrated, featuring a comparison between the analytical linear approximation (green dashed line) and the numerical calculations (red curve). The vertical lines represent the eigenvalues of the linear modes, with the bold vertical line specifically highlighting the eigenvalue of the linear edge mode from which the nonlinear bifurcation originates.
}}
}
\label{figS8}
\end{figure}

{{

\section{Low-power limit for the edge soliton family in quasicrystal array}\label{sec9}

In this section, we present an approximation for the $U(b)$ dependence in the linear limit, i.e., close to the point, where soliton bifurcates from linear edge mode. To achieve this, we follow the method outlined for quasi-periodic nonlinear systems in~\cite{Konotop2024} and assume that the soliton field $\psi(\br,z)$ can be expanded as $\psi=e^{ibz} \sum_{j=0}^N a_j(z) w_j(\br)$, where the modal amplitudes $a_j$ of different linear eigenmodes of the system $w_j(\br)$ solve the system
\[
i \frac{da_j}{dz} = (b-b_j)a_j + \sum_{k,l,m=0}^N \gamma_{jklm} a^*_k a_l a_m.
\]
Here, $b_j$ are the propagation constants of linear mode with index $j$, and the nonlinear coefficients are given by
\[
\gamma_{jklm} = 
\begin{cases}
\int w_jw_k w_l^*w_m^* d^2\br, \quad &\textrm{for} \quad 1\leq j,k,l,m\leq N,\\
0,\quad &\textrm{otherwise}
\end{cases}
\]
$w_j$ are the functions defining the transverse profile of linear mode $j$, and the asterisk denotes complex conjugation. The expansion coefficients are normalized such that $\sum_{j=1}^N|a_j|^2=U$. It can be demonstrated (see~\cite{Konotop2024}) that the nonlinear coefficients involving the interaction (or ``hopping'') of three or four different modes are significantly smaller than either $\gamma_{jkjk}$ or $\gamma_{jjjk}$. Thus, all terms involving $\gamma_{jklm}$, where three or all four indices are distinct, are neglected. Under this approximation, the above system of equations reduces to a form:
\begin{align*}
i \frac{da_j}{dz} = &(b-b_j)a_j + \gamma_{jjjj} |a_j|^2a_j  \\
+&\sum_{k\neq j}\left[ \gamma_{jkjk}(2|a_j|^2a_k+a_j^2a_k^*) + \gamma_{kkkj} |a_k|^2a_k\right]
\end{align*}
The simplest approximate solution to this system, often referred to as a ``monomer'' solution, occurs when only one mode (say, the $j$-th mode) is excited, such that $a_n = \sqrt{U}$. Namely this scenario is realized when we excite the edge state within our truncated quasiperiodic array. Then the power for this solution is given by
\begin{align*}
U(b)=\frac{b-b_n}{\gamma_{nnnn}}.
\end{align*}
Here, $\gamma_{nnnn}=\int |w_n|^4 d^2\br$. The corresponding linear dependence is plotted in Fig.~\ref{figS8} as a green dashed line together with the exact $U(b)$ soliton family (red curve). One can see that near the bifurcation point, the analytical prediction very accurately describes the slope of the exact $U(b)$ dependence. However, this alignment diverges when soliton coupling with bulk linear modes (see vertical lines indicating eigenvalues of these modes) occurs, resulting in a sharply increasing slope of the red curve. Notably, this soliton family does not exhibit any power threshold, as it bifurcates from the linear edge mode, whose existence is numerically validated in the main text of the Letter, The complete nonlinear family for the quasiperiodic array is depicted in Fig.~3 of the main text. 

}}




\begin{thebibliography}{99}



\bibitem{Janot-12}
C.~Janot, Quasicrystals, A Primer. Clarendon Press, Oxford (2012).

\bibitem{Steurer-18}
W.~Steurer,
Quasicrystals: What do we know? What do we want to know? What can we know?,
Acta Crystallogr. A {\bf 574}, 1--11 (2018).

\bibitem{Shechtman-84}
D.~Shechtman, I.~Blech, D.~Gratias, J.W.~Cahn,
Metallic Phase with Long-Range Orientational Order and No Translational Symmetry,
Phys. Rev. Lett. {\bf 53}, 1951--1953 (1984).



\bibitem{CollinsWitte-17}
L.C.~Collins, T.G.~Witte, R.~Silverman, D.B.~Green, K.K.~Gomes,
Imaging quasiperiodic electronic states in a synthetic penrose tiling,
Nature Communications {\bf 8}(1), 15961 (2017).

\bibitem{KempkesSlot-21}
S.N.~Kempkes, M.R.~Slot, S.E.~Freeney, S.J.M.~Zevenhuizen, D.~Vanmaekelbergh, I.~Swart, C.M.~Smith,
Design and characterization of electrons in a fractal geometry,
Nature Physics {\bf 15}(2), 127--131 (2019).



\bibitem{Steurer-04}
W.~Steurer, Twenty years of structure research on quasicrystals.
Part I. Pentagonal, octagonal, decagonal and dodecagonal
quasicrystals. Z. Kristallogr. Cryst. Mater. {\bf 219}, 391–446 (2004).

\bibitem{KamiyaTakeuchi-18}
K.~Kamiya, T.~Takeuchi, N.~Kabeya, N.~Wada, T.~Ishimasa, A.~Ochiai, K.~Deguchi, K.~Imura, and N.K.~Sato,
Discovery of superconductivity in quasicrystal,
Nat. Commun. {\bf 9}, 1--8 (2018).

\bibitem{AhnMoon18}
S.J.~Ahn, P.~Moon, T.-H.~Kim, H.-W.~Kim, H.-C.~Shin, E.H.~Kim, H.W.~Cha, S.-J.~Kahng, P.~Kim, M.~Koshino, Y.-W.~Son, C.-W.~Yang, J.R.~Ahn,
Dirac electrons in a dodecagonal graphene quasicrystal,
Science {\bf 361}, 782--786 (2018).

\bibitem{YaoaWanga-18}
W.~Yaoa, E.~Wanga, C.~Baoa, Y.~Zhangc, K.~Zhanga, K.~Baoc, C.K.~Chanc, C.~Chend, J.~Avilad, M.~C.~Asensiod, J.~Zhuc, and S.~Zhou,
Quasicrystalline $30^{\circ}$ twisted bilayer graphene as an
incommensurate superlattice with strong interlayer coupling,
Proc. Natl Acad. Sci. USA {\bf 115}, 6928--6933 (2018).



\bibitem{VerreAntosiewicz-14}
R.~Verre, T.J.~Antosiewicz, M.~Svedendahl, K.~Lodewijks, T.~Shegai, M.~Kä\"{a}ll,
Quasi-isotropic surface plasmon polariton generation through near-field coupling to a penrose pattern of silver nanoparticles,
ACS Nano {\bf 8}(9), 9286--9294 (2014).



\bibitem{WatanabeBhat-21}
S.~Watanabe, V.S.~Bhat, K.~Baumgaertl, M.~Hamdi, D.~Grundler,
Direct observation of multiband transport in magnonic penrose quasicrystals via broadband and phase-resolved spectroscopy,
Science Advances {\bf 7}(35), 3771 (2021).



\bibitem{SchreiberHodgman-15}
M.~Schreiber, S.S.~Hodgman, P.~Bordia, H.P.~L\"{u}schen, M.H.~Fischer, R.~Vosk, E.~Altman, U.~Schneider, I.~Bloch,
Observation of many-body localization of interacting fermions in a quasirandom optical lattice,
Science {\bf 349}(6250), 842--845 (2015).

\bibitem{ViebahnSbroscia-19}
K.~Viebahn, M.~Sbroscia, E.~Carter, J.-C.~Yu, U.~Schneider,
Matter-wave diffraction from a quasicrystalline optical lattice,
Phys. Rev. Lett. {\bf 122}, 110404 (2019).

\bibitem{SbrosciaViebahn-20}
M.~Sbroscia, K.~Viebahn, E.~Carter, J.C.~Yu, A.~Gaunt, and U.~Schneider,
Observing localization in a 2D quasicrystalline optical lattice,
Phys. Rev. Lett. {\bf 125}, 200604 (2020).

\bibitem{SanchezLewenstein-10}
L.~Sanchez-Palencia and M.~Lewenstein,
Disordered quantum gases under control,
Nat. Phys. {\bf 6}, 87--95 (2010).


\bibitem{TaneseGurevich-14}
D.~Tanese, E.~Gurevich, F.~Baboux, T.~Jacqmin, A.~Lema\^{i}tre, E.~Galopin, I.~Sagnes, A.~Amo, J.~Bloch, E.~Akkermans,
Fractal energy spectrum of a polariton gas in a fibonacci quasiperiodic potential,
Phys. Rev. Lett. {\bf 112}, 146404 (2014).

\bibitem{BabouxLevy-17}
F.~Baboux, E.~Levy, A.~Lema\^{i}tre, C.~G\'{o}mez, E.~Galopin, L.~Le~Gratiet, I.~Sagnes, A.~Amo, J.~Bloch, E.~Akkermans,
Measuring topological invariants from generalized edge states in polaritonic quasicrystals,
Phys. Rev. B {\bf 95}, 161114(R) (2017).

\bibitem{GoblotStrkalj-20}
V.~Goblot, A.~\v{S}trkalj, N.~Pernet, J.L.~Lado, C.~Dorow, A.~Lema\^{i}tre, L.~Le~Gratiet, A.~Harouri, I.~Sagnes, S.~Ravets, A.~Amo, J.~Bloch, O.~Zilberberg,
Emergence of criticality through a cascade of delocalization transitions in quasiperiodic chains,
Nature Physics {\bf 16}(8), 832--836 (2020).



\bibitem{VardenyNahata-13}
Z.V.~Vardeny, A.~Nahata, A.~Agrawal,
Optics of photonic quasicrystals,
Nature Photonics {\bf 7}(3), 177--187 (2013).

\bibitem{XuWang-21}
X.-Y.~Xu, Wang, D.-Y.~Chen, C.M.~Smith, X.-M.~Jin,
Quantum transport in fractal networks,
Nature Photonics {\bf 15}(9), 703--710 (2021).

\bibitem{WangFuKonotop-24}
P.~Wang, Q.~Fu, V.V.~Konotop, Y.V.~Kartashov, F.~Ye,
Observation of localization of light in linear photonic quasicrystals with diverse rotational symmetries,
Nature Photonics {\bf 18}, 224--229 (2024) (2024).




\bibitem{MatsuiAgrawal-07}
T.~Matsui, A.~Agrawal, A.~Nahata, Z.V.~Vardeny,
Transmission resonances through aperiodic arrays of subwavelength apertures,
Nature {\bf 446}(7135), 517--521 (2007).


\bibitem{LeviRechtsman-11}
L.~Levi, M.~Rechtsman, B.~Freedman, T.~Schwartz, O.~Manela, M.~Segev,
Disorder-enhanced transport in photonic quasicrystals,
Science {\bf 332}, 1541--1544 (2011).

\bibitem{Wiersma-13}
D.S.~Wiersma,
Disordered photonics,
Nat. Photon. {\bf 7}, 188-196 (2013).

\bibitem{SegevSilberberg-13}
M.~Segev, Y.~Silberberg, and D.N.~Christodoulides,
Anderson localization of light,
Nat. Photon. {\bf 7}, 197--204 (2013).


\bibitem{BandresRechtsman-16}
M.A.~Bandres, M.C.~Rechtsman, M.~Segev,
Topological photonic quasicrystals: Fractal topological spectrum and protected transport,
Phys. Rev. X {\bf 6}, 011016 (2016).

\bibitem{ZhouZhang-19}
D.~Zhou, L.~Zhang, and X.~Mao,
Topological Boundary Floppy Modes in Quasicrystals,
Phys. Rev. X {\bf 9}, 021054 (2019).




\bibitem{WangZheng-20}
P.~Wang, Y.~Zheng, X.~Chen, C.~Huang, Y.V.~Kartashov, L.~Torner, V.V.~Konotop, and F.~Ye,
Localization and delocalization of light in photonic moir\'{e} lattices,
Nature (London) {\bf 577}, 42 (2020).

\bibitem{HuangYe-16}
C.~Huang, F.~Ye, X.~Chen, Y.V.~Kartashov, V.V.~Konotop, and L.~Torner,
Localization-delocalization wavepacket transition in Pythagorean aperiodic potentials,
Sci. Rep. {\bf 6}, 32546 (2016).





\bibitem{FreedmanBartal-06}
B.~Freedman, G.~Bartal, M.~Segev, R.~Lifshitz, D.N.~Christodoulides, J.W.~Fleischer,
Wave and defect dynamics in nonlinear photonic quasicrystals,
Nature {\bf 440}(7088), 1166--1169 (2006).

\bibitem{FreedmanLifshitz-07}
B.~Freedman, R.~Lifshitz, J.W.~Fleischer, M.~Segev,
Phason dynamics in nonlinear photonic quasicrystals,
Nature Materials {\bf 6}(10), 776--781 (2007).




\bibitem{ClausenKivshar-99}
C.A.B.~Clausen, Y.S.~Kivshar, O.~Bang, and P.L.~Christiansen,
Quasiperiodic envelope solitons,
Phys. Rev. Lett. {\bf 83}, 4740--4743 (1999).

\bibitem{AblowitzIlan-06}
M.J.~Ablowitz, B.~Ilan, E.~Schonbrun, and R.~Piestun,
Solitons in two-dimensional lattices possessing defects, dislocations, and quasicrystal structures,
Phys. Rev. E {\bf 74}, 035601(R) (2006).

\bibitem{LawSaxena-10}
K.J.H.~Law, A.~Saxena, P.G.~Kevrekidis, and A.R.~Bishop,
Stable structures with high topological charge in nonlinear photonic quasicrystals,
Phys. Rev. A {\bf 82}, 035802 (2010).

\bibitem{AblowitzAntar-12}
M.J.~Ablowitz, N.~Antar, İ.~Bakırta\c{s}, and B.~Ilan,
Vortex and dipole solitons in complex two-dimensional nonlinear lattices,
Phys. Rev. A {\bf 86}, 033804 (2012).

\bibitem{FuWang-20}
Q.~Fu, P.~Wang, C.~Huang, Y.V.~Kartashov, L.~Torner, V.V.~Konotop, and F.~Ye,
Optical soliton formation controlled by angle twisting in photonic moir\'{e} lattices,
Nat. Photonics {\bf 14}, 663 (2020).

\bibitem{KartashovYe-21}
Y.V.~Kartashov, F.~Ye, V.V.~Konotop, and L.~Torner,
Multifrequency Solitons in Commensurate-Incommensurate Photonic Moir\'{e} Lattices,
Phys. Rev. Lett. {\bf 127}, 163902 (2021).

\bibitem{HuangDong-21}
C.~Huang, L.~Dong, H.~Deng, X.~Zhang, and P.~Gao,
Fundamental and vortex gap solitons in quasiperiodic photonic lattices,
Optics Letters 46, 22, 5691--5694 (2021)




\bibitem{Penrose-79}
R.~Penrose,
Pentaplexity: A class of non-periodic tilings of the plane,
The Mathematical Intelligencer {\bf 2}, 32--37 (1979).











\bibitem{Konotop2024}
V.V.~Konotop,
Dimers and discrete breathers in Bose-Einstein condensates in a quasi-periodic potential,
Phys. Rev. Research {\bf 6}, 033113 (2024).


\bibitem{VakhitovKolokolov-73}
N.G.~Vakhitov and A.A.~Kolokolov,
Stationary solutions of the wave equation in a medium with nonlinearity saturation,
Radiophys. Quantum Electron. {\bf 16}, 783 (1973).





\bibitem{Berge1998}
L.~Berg\'{e},
Wave collapse in physics: principles and applications to light and plasma waves,
Phys. Rep. \textbf{303}, 259 (1998).

\bibitem{AcevesTuritsyn1994}
A.B.~Aceves, C.~De~Angelis, A.M.~Rubenchik, S.K.~Turitsyn,
Multidimensional solitons in fiber arrays,
Opt. Lett. \textbf{19}, 329 (1994).

\bibitem{AcevesTuritsyn1995}
A.B.~Aceves, G.G.~Luther, C.~De~Angelis, A.M.~Rubenchik, S.K.~Turitsyn,
Energy Localization in Nonlinear Fiber Arrays: Collapse-Effect Compressor,
Phys. Rev. Lett. \textbf{75}, 73 (1995).

\bibitem{KivsharPelinovsky2000}
Y.S.~Kivshar, D.E.~Pelinovsky,
Self-focusing and transverse instabilities of solitary waves,
Phys. Rep. \textbf{331}, 117 (2000).

\bibitem{Kartashov-19}
Y.V.~Kartashov, G.E.~Astrakharchik, B.A.~Malomed, L.~Torner,
Frontiers in multidimensional self-trapping of nonlinear fields and matter,
Nat. Rev. Phys. \textbf{1}, 185 (2019). 

\bibitem{Malomed2024}
B.A.~Malomed,
Multidimensional soliton systems,
Advances in Physics X \textbf{9}, 2301592 (2024).

\bibitem{ZhangKartashov2024}
H.~Zhong, V.O.~Kompanets, Y.~Zhang, Y.V.~Kartashov, M.~Cao, Y.~Li, S.A.~Zhuravitskii, N.N.~Skryabin, I.V.~Dyakonov, A.A.~Kalinkin, S.P.~Kulik, S.V.~Chekalin, and V.N.~Zadkov,
Observation of nonlinear fractal higher order topological insulator,
Light: Science \& Applications {\bf 3}, 264 (2024).





\bibitem{Ferrando2005c}
A.~Ferrando,
Discrete-symmetry vortices as angular Bloch modes,
Phys. Rev. E, {\bf 72}(3), 036612 (2005). 

\bibitem{hamermesh64}
M.~Hamermesh,
Group theory and its application to physical problems,
(First ed.) Reading, Massachusetts: Addison-Wesley (1964). 

\bibitem{Asboth2016}
J.K.~Asb{\'o}th, L.~Oroszl{\'a}ny, and A.~P{\'a}lyi,
A Short Course on Topological Insulators,
Cham: Springer International Publishing (2016).



\bibitem{Ferrando2017}
A.~Ferrando,
\newblock Nonlinear plasmonic amplification via dissipative soliton-plasmon resonances,
\newblock {Phys. Rev. A}, {\bf 95}, 013816 (2017).






\bibitem{SulemSulem1999}
C.~Sulem and P.-L.~Sulem,
The Nonlinear Schr\"{o}dinger Equation: Self-Focusing and Wave Collapse (Springer, New York, 1999).


\bibitem{KartashovMalomed2011}
Y.V.~Kartashov, B.A.~Malomed, L.~Torner,
Solitons in nonlinear lattices,
Reviews of Modern Physics {\bf 83}(1), 247--305 (2011).

\bibitem{Fibich2015}
G.~Fibich,
The Nonlinear Schr\"{o}dinger Equation: Singular Solutions and Optical Collapse (Springer, Heidelberg, 2015).







\bibitem{Yang2010}
J.~Yang,
in Nonlinear Waves in Integrable and Nonintegrable
Systems, Mathematical Modeling and Computation (SIAM,
Philadelphia, 2010).






\bibitem{PelinovskyKevrekidisFrantzeskakis2005}
D.E.~Pelinovsky, P.G.~Kevrekidis, and D.J.~Frantzeskakis,
Stability of discrete solitons in nonlinear Schr\"{o}dinger lattices,
Physica D: Nonlinear Phenomena {\bf 212}, 1--2, 1, 1--19 (2005).

\bibitem{KartashovMalomed2009}
Y.V.~Kartashov, B.A.~Malomed, V.A.~Vysloukh, and L.~Torner,
Two-dimensional solitons in nonlinear lattices,
Opt. Lett. {\bf 34}, 770 (2009).

\bibitem{ChongPelinovsky2011}
C.~Chong and D.~Pelinovsky,
Variational approximations of bifurcations of asymmetric solitons in cubic-quintic nonlinear Schr\"{o}dinger lattices,
Discrete and Continuous Dynamical Systems - S {\bf 4}(5) 1019--1031 (2011).













\end{thebibliography}
\end{document}